\documentclass[%
preprint,
amsmath,amssymb,
prb,
]{revtex4-1}

\pdfoutput=1
\usepackage{graphicx}
\usepackage{dcolumn}
\usepackage{bm}

\begin{document}

\title[The Einstein - de Haas effect at radio frequencies in yttrium iron garnet]{The Einstein - de Haas effect at radio frequencies in and near magnetic equilibrium}

\author{K. Mori$^1$}
\author{M.G. Dunsmore$^1$}%
\author{J.E. Losby$^{1,2}$}%
\author{D.M. Jenson$^1$}%
\author{M. Belov$^2$}%
\author{M.R. Freeman$^1$}

\email{mark.freeman@ualberta.ca}

\affiliation{%
$^1$ Department of Physics, University of Alberta, Edmonton, Alberta T6G 2E1, Canada
}%

\affiliation{%
$^2$ Nanotechnology Research Centre, National Research Council
of Canada, Edmonton, Alberta T6G 2M9, Canada
}%


\begin{abstract}
The Einstein-de Haas (EdH) effect and its reciprocal the Barnett effect are fundamental to magnetism and uniquely yield measures of the ratio of magnetic moment to total angular momentum.  These effects, small and generally difficult to observe, are enjoying a resurgence of interest as contemporary techniques enable new approaches to their study.  The high mechanical resonance frequencies in nanomechanical systems offer a tremendous advantage for the observation of EdH torques in particular. At radio frequencies, the EdH effect can become comparable to or even exceed in magnitude conventional cross-product magnetic torques.  In addition, the RF-EdH torque is expected to be phase-shifted $90^{\circ}$ relative to cross-product torques, provided the magnetic system remains in quasi-static equilibrium, enabling separation in quadratures when both sources of torque are operative.  Radio frequency EdH measurements are demonstrated through the full hysteresis range of micrometer scale, monocrystalline yttrium iron garnet (YIG) disks.  Equilibrium behavior is observed in the vortex state at low bias field.  Barkhausen-like features emerge in the in-plane EdH torque at higher fields in the vortex state, revealing magnetic disorder too weak to be visible through the in-plane cross-product torque.  Beyond vortex annihilation, peaks arise in the EdH torque versus bias field, and these together with their phase signatures indicate additional utility of the Einstein-de Haas effect for the study of RF-driven spin dynamics.  
\end{abstract}


\maketitle

\pagebreak
\section{\label{SecI}Introduction}

Einstein-de Haas (EdH) measurements on magnetic systems, historically, have been challenging \cite{Einstein1915,Stewart1918,Barnett1935,Scott1962,Galison1987} owing to the small magnitudes of EdH torques in relation to conventional, cross-product of field and moment vector magnetic torques.  EdH measurements involve application of an alternating magnetic field \textit{parallel} to a mechanical torsion axis in order to probe the intrinsic, microscopic coupling of magnetic moment and mechanical angular momentum in the sample.  The angular momentum variation accompanying field-driven changes in net moment is governed by the material's magnetomechanical ratio, $g\prime$ \cite{Kittel1949}.  Determinations of $g\prime\thinspace$ have been the primary motivation for EdH studies.  There is also an opportunity to exploit the EdH effect to measure AC susceptibilities along the direction of an applied AC magnetic field.  

A conventional magnetic torque experiment \cite{Cullity2008}, by contrast intrinsically measures magnetic anisotropies, by sensing the magnetic potential energy variation versus angle for a magnetic object in an external applied field.  Anisotropies can arise from specimen shape, composition, and microstructure via dipolar, spin-orbit, exchange, and interfacial couplings.  When the anisotropy is known, the cross-product torque is useful for the determination of magnetic moment (torque magnetometry).  The ratio of net magnetic moment, $m$, to net angular momentum, $J$, in a magnetic material is given in terms of $g\prime$ by 

\begin{equation}
\frac{m}{J} = g\prime \frac{e}{2 m_\textrm{e}} \equiv \gamma\prime
\label{eq:one}
\end{equation}

\noindent where $e$ and $m_\textrm{e}$ are the electron charge and mass.  We define this quantity as the spin-mechanical gyromagnetic ratio of the material, $\gamma\prime$, to position mechanical torque experiments on a parallel footing with spin precession measurements.  Simultaneous EdH and cross-product torque measurements represent an opportunity to determine $\gamma\prime$ directly, via ratios of the mechanical signal amplitudes.  

The miniaturization of torque sensors that began late in the 20th century with silicon micromachining \cite{Kleiman1985} has created new opportunities for study of the EdH effect.  An elegant determination of $g\prime$ for a Permalloy thin film was reported in 2006  by Wallis \textit{et al.} \cite{Wallis2006}, who deposited the sample on a silicon microcantilever and measured at the 13 kHz fundamental flexural resonance of that device.  The Einstein - de Haas effect in a 23 kHz YIG cantilever has been used to detect angular momentum pumping driven by the spin Seebeck effect \cite{Harii2019}.  Finally, the physics of the Barnett effect has been elucidated recently through inductively detected electron and nuclear spin resonance experiments, with and without the detection coil in the reference frame of the rotating specimen \cite{Arabgol2019,Chudo2015}.  

The present work draws attention to distinguishing features of EdH torques that emerge as EdH experiments are miniaturized further, both in the quasi-static regime where the magnetic system remains at or near instantaneous equilibrium with the total applied field, and as spin dynamics begin to emerge.  

\begin{figure*}
\includegraphics[width=16.4cm,keepaspectratio]{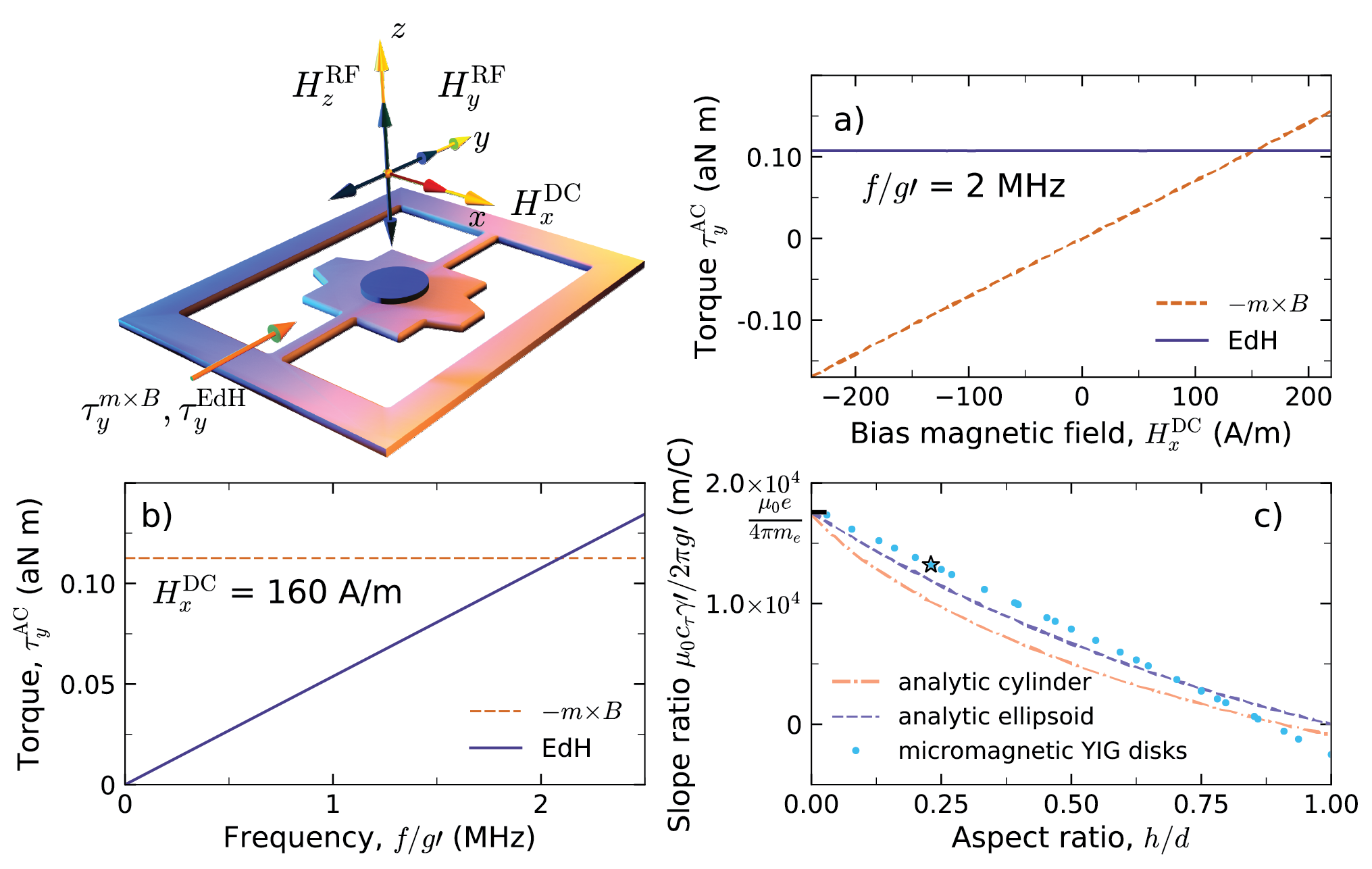}
\caption{\label{Fig1}Conceptual overview of simultaneous cross-product and EdH torque measurements on a micromagnetic disk supported by a mechanical torsion resonator. The RF and DC magnetic field geometry is illustrated at the upper left.  $H_y^{RF}$ drives the EdH torque, $\tau_{y}^\textrm{ EdH}$, while $H_z^{RF}$ drives the cross-product torque, $\tau_y^{m\times B}$. The DC field along $x$ is variable, and swept for hysteresis measurements. a) Simulated torques for a 2.1 $\mu$m diameter by 490 nm thick YIG disk with a magnetic vortex spin texture at low bias fields, showing the cross-product torque linear proportionality to field, and the EdH torque field strength-independence.   b) Simulated torques as a function of frequency, showing the EdH torque strength proportionality to frequency, and the cross-product torque frequency-independence.  c) The ratio of the linear slopes as in panels b) and c), as a function of sample geometry.  The physical constant governing the overall scale of torque slope ratios is the inverse electron gyromagnetic ratio.  Lines: analytical results for cylinders and ellipsoids of revolution, assuming uniform magnetization.  Symbols: a collection of micromagnetic simulations of YIG disks, including cubic anisotropy with an easy axis oriented along $\hat{z}$. The star corresponds to the aspect ratio used for panels a) and b).  All torque amplitudes in this figure are calculated for RF drive field strengths of 80 A/m RMS.}
\end{figure*}

\section{\label{SecII}Simultaneous Einstein-de Haas and cross-product\\ AC magnetic torques: conceptual framework}

 A quantitative sense of the magnitude of the quasi-static RF-EdH effect for standard specimens is obtained by considering mechanical torque measurements on a disk of soft ferromagnetic material, in the configuration illustrated in the upper left panel of Fig.~\ref{Fig1}.  The sample material is assumed to be magnetically isotropic with a constant volume susceptibility, $\chi$, at low fields.   The shape has magnetometric demagnetizing factors $N_z$ and $N_r = (1-N_z)/2$ in the axial and radial directions \cite{Joseph1966, Beleggia2006b}.  Application of RF and DC magnetic fields, $H_{z}^\textrm{RF}\sin(\omega t)$ and $H_x^\textrm{DC}$ to the disk, both assumed small enough to remain in the constant susceptibility regime, and taking $B = \mu_0 H$ for all external applied fields, yields the RF cross-product torque


\begin{equation}
\begin{split}
\tau_{y}^{m\times B} & = -(\chi V (1 - N_r)H_x^\textrm{DC})(\mu_0 H_z^\textrm{RF} \sin(\omega t)) \\
& \indent + (\chi V (1 - N_z)H_z^\textrm{RF} \sin(\omega t))(\mu_0 H_x^\textrm{DC}) \\ 
& = -\mu_0 (N_z - N_r) \chi V H_x^\textrm{DC} H_z^\textrm{RF}\sin(\omega t).
\end{split}
\label{eq:two}
\end{equation}

\noindent  Equation~\ref{eq:two} represents the resultant $y$-torque, after the contribution from the moment along $x$ multiplied by the RF field along $z$ (first term) is reduced by the induced RF moment along $z$ multiplied by the DC field along $x$ (the second term is significant outside the very thin film limit where $N_z \approx 1$).  This same RF drive will generate an EdH mechanical torque along $z$ (not measurable with the devices used here), according to

\begin{equation}
\begin{split}
\tau_{z}^\textrm{EdH} & = -\frac{dJ_z}{dt}\\
& = -\frac{2 m_\textrm{e}}{e g\prime} (1-N_z) \chi V \omega H_z^\textrm{RF} \cos(\omega t),
\end{split}
\label{eq:three}
\end{equation}

\noindent where the minus sign comes from total angular momentum conservation of the combined magnetic/mechanical system in the absence of external torques. 

The EdH torque of Eq.~\ref{eq:three} reflects the RF magnetic susceptibility along $\hat{z}$.  A linear increase with frequency of the relative amplitudes of EdH and cross-product torques is expected, along with a quadrature phase relationship between the two signals.  Both consequences arise from the single time derivative of the mechanical angular momentum underlying the EdH torque. 

The current experiment has a single torque sensing axis and therefore a separate RF drive field, $H_y^\textrm{RF}\sin(\omega t)$, is used to excite an in-plane EdH torque,

\begin{equation}
\begin{split}
\tau_{y}^\textrm{EdH} & = -\frac{dJ_y}{dt}\\
& = -\frac{2 m_\textrm{e}}{e g\prime} (1-N_r) \chi V \omega H_y^\textrm{RF} \cos(\omega t),
\end{split}
\label{eq:four}
\end{equation}

We turn to numerical simulation of the torques with MuMax3 \cite{Vansteenkiste2014} to account for the influence of nonuniform spin textures, magnetocrystalline anisotropy, and other modifications beyond the simplest scenario.  Results from a micromagnetic simulation of RF cross-product and EdH torques for a yttrium iron garnet disk of height-to-diameter aspect ratio 0.23 and in the vortex state are plotted in Figs.~\ref{Fig1}a and~\ref{Fig1}b, respectively, versus low bias fields, $H_x$ (omitting the superscript DC from this point forward), for fixed frequency and versus frequency, $f$, for fixed low bias field.  These graphs emphasize the take-away from equations~\ref{eq:two} and~\ref{eq:three} that the torques are linearly proportional respectively to $H_x$ and $f = \omega/2\pi$, and independent of the other parameter, when the system remains quasi-statically in magnetic equilibrium.  The simulation predicts crossover of the EdH and cross-product torque magnitudes at $f/g\prime = 2$ MHz and $H_x = 160$ A/m, as a specific example.  

An estimation of the relative torques for arbitrary shapes can be obtained from a plot of the ratio of these two slopes, which in the linear regime can be written

\begin{equation}
\frac{d\tau_{y}^{m\times B}/dH_x}{d\tau_y^\textrm{EdH}/df} = c_\tau \frac{\mu_0 \gamma\prime}{2\pi},
\label{eq:five}
\end{equation}

\noindent where $c_\tau$ is a sample-specific dimensionless constant accounting for anisotropy.  In the analytic case presented above, $c_\tau =  (N_z - N_r)/(1 - N_r) = (1 - 3N_r)/(1 - N_r).$  

When the two RF drives are of the same strength, the predicted ratio of the two torque amplitudes is 

\begin{equation}
\frac{\tau_{y}^{m\times B}}{\tau_{y}^\textrm{EdH}} = c_\tau \frac{\gamma\prime \mu_0}{2\pi} \frac{H_x}{f}. \label{eq:six}
\end{equation} 

\noindent  An analog of the expression for Larmor precession arises for conditions where the torque magnitudes are equal and the left-hand side therefore equals unity.  In the Larmor case, the frequency is governed by a ratio of magnetic torque to spin angular momentum, governed by the relevant spectroscopic splitting g-factor and generally also with at least a small modification by a magnetic anisotropy term \cite{Christensen2017}.  

Fig.~\ref{Fig1}c summarizes these results, and presents a quantity that is independent of the material-specific $g\prime$ by showing $(c_\tau \mu_0 \gamma\prime/2\pi g\prime)$ versus the height-over-diameter aspect ratio, $h/d$.   (The division by $g\prime$ cancels the $g\prime$ built-in to $\gamma\prime$.)  The scale of torque slope ratios is anchored by $4\pi m_\textrm{e}/(\mu_0 e) = 17588\thinspace$m/C, for fields and frequencies measured in A/m and Hz.  To describe any particular sample, the 17588 m/C value is multiplied by the corresponding $c_\tau$ accounting for the aggregate anisotropy.  The torque slope ratio decreases as the sample becomes more isotropic, for example as a thin disk becomes taller and the cross-product torque decreases.  In the case of the disk, this decrease of cross-product torque is slightly offset by a decrease of the in-plane susceptibility and hence of the EdH torque, as the disk becomes taller.  As numerical examples in the limit of a thin disk with $h/d \sim 0$, and with $f/g\prime = 2\thinspace$MHz, the cross-product torque will equal the EdH torque when $H_x \equiv H_\textrm{crossover} = 4 \times 10^6\thinspace$s$^{-1}$/($2\times17588$ m/C) = 114 A/m.  This is a very achievable condition for routine micromechanical magnetometry.  In comparison, note that for the mechanical frequency of Ref. Wallis (13 kHz), we would have $H_\textrm{crossover} = 0.37\thinspace$A/m (approximately 1/100 of Earth's field), and for that of the original EdH experiment (50 Hz), it would be about 1/25,000 of Earth's field.  

Results from a series of micromagnetic simulations for yttrium iron garnet cylinders of various aspect ratios, all in the magnetic vortex spin texture, are shown by the closed symbols in Fig.~\ref{Fig1}c.  The cubic anisotropy is oriented to have an easy axis along $z$, which is also the axis of the vortex core.  The open star demarcates the aspect ratio corresponding to the simulation results shown in Figs.~\ref{Fig1}a,b.  Again, to yield the same ($y$) component of torque, the AC drive field directions for EdH, cross-product torques in the calculation are along $\hat{y}, \hat{z}$, respectively.  Analytical curves representing the expected ratio of cross-product and Einstein - de Haas AC torque magnitudes are presented also, for cylinders and ellipsoids of revolution of any soft ferromagnetic material having an isotropic volume susceptibility.  The dashed curve for the ellipsoid of revolution has its zero crossing at aspect ratio = 1 (sphere), where the cross-product torque vanishes.  The corresponding aspect ratio where the radial and axial demagnetizing factors for the cylinder are equal (where its axial and radial demagnetizing factors are equal) is $h/d = 0.9$.  At low aspect ratio, the more rapid initial decrease of the torque slope ratio for ellipsoids versus cylinders reflects the more rapid increase of the radial demagnetizing factor for the cylinders, as the height increases.  

Qualitatively, the torque slope ratio from the simulations at low aspect ratio is closer to the analytical result for the ellipsoids of revolution; the vortex texture helps sustain the ease of magnetizing in-plane, initially as the height increases.  The taller YIG disks show a net-zero cross-product torque at an aspect ratio close to that of the analytical cylinders.  The vortex texture and cubic anisotropy conspire overall to yield a more linear decrease of the torque slope ratio versus aspect ratio, in comparison to either of the analytical results.  The contributions of the spin texture and anisotropy are not fully independent, in modifying the torques over the simple analytical case.  For the sample orientation in this study, having a magnetocrystalline easy axis parallel to the cylinder axis, negative $K_c$ also helps to stabilize the vortex texture in taller cylinders.  The coupling of these effects underscores the importance of performing micromagnetic simulations of the torques.    

\section{\label{SecIII}Experimental}

\begin{figure}[htb!]
\includegraphics{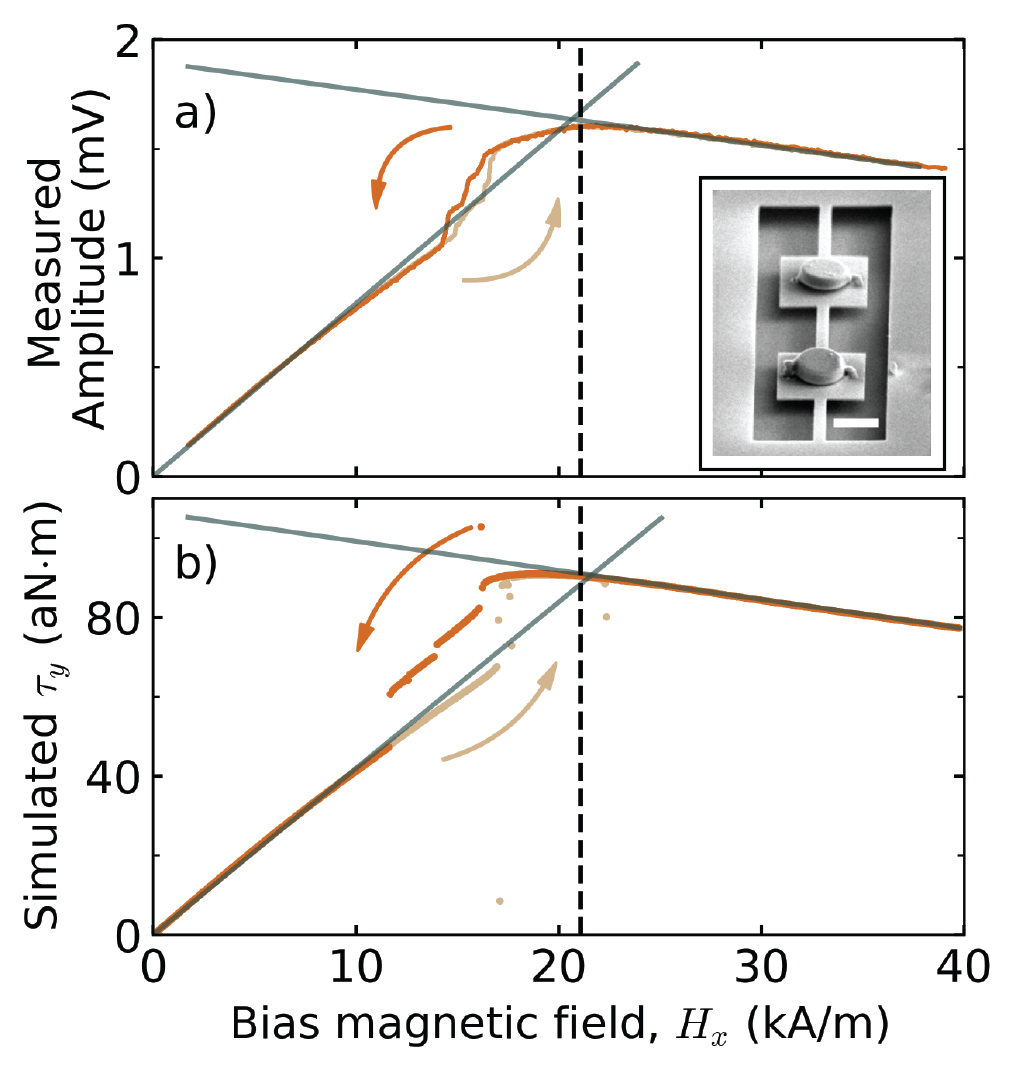}
\caption{\label{Fig2}Cross-product torque hysteresis loops, measured and simulated.  a) The experimental cross-product torque hysteresis of the double disk sample, driven with an RF field along $\hat{z}$ at RMS amplitude 247$\pm$7 A/m.  The conversion factor between lock-in signal magnitude and torque, obtained by thermomechanical calibration, is  $ 41 \pm 1$ aN m / mV.   A scanning electron micrograph of the sample is shown in the inset (scale bar = 2$\thinspace \mu$m).  b) Cross-product torque hysteresis from micromagnetic simulation of a single disk of 2.0 $\mu$m diameter and 0.46 $\mu$m thickness, subject to the same RF drive amplitude used for the measurements.  The torque is multiplied by two to correspond to a net magnetic volume similar to that of the sample.  The two disks of the sample are far enough apart to neglect the dipole coupling between them.  In a) and b) the solid lines show linear fits to the low field (vortex state) and high field (quasi-uniform texture) torques.  Where these lines intersect is a sensitive indicator of aspect ratio, here at approximately 21 kA/m as indicated by the vertical dashed line.  The saturation moment from the measurements, as determined by comparing the peak experimental and simulated torques, is $m_\textrm{s} = M_\textrm{s}V = (3.0\pm0.1)\times 10^{-13}$ A$\cdot$m$^{2}$.}
\end{figure}

For comparative measurements of EdH and cross-product torques, an array of mechanical sensors is pre-fabricated through standard silicon-on-insulator processing using electron beam lithography and etching \cite{Diao2013}.  YIG disks of design thickness/diameter aspect ratio $h/d \sim 0.3\thinspace$ are cut with a Ga$^{+}$ ion beam from a thin polished wedge, transferred by nanomanipulation onto the torque sensors, and tack-welded in place with a carbonaceous ion beam-induced deposit. Monocrystalline YIG disks are a favourable choice for the demonstration: they demagnetize into a vortex state yielding zero cross-product torque at approximately zero DC bias field; and their low-field magnetizing curves are not complicated by pinning effects.  The as-fabricated \textit{magnetic} volumes and disk aspect ratios are somewhat lower than the physical dimensions on account of magnetically-dead layers that are created by Ga beam damage to a depth of at least 50 nm under surfaces exposed to the ion flux \cite{Fraser2010}.  The results reported below are from the double paddle device shown in the inset to Fig.~\ref{Fig2}, supporting two disks that are equivalent to within the fabrication uncertainty.  The equations of Sec.~\ref{SecII} apply without modification to this structure.  The double paddle device had the best signal transduction out of several structures that were fabricated.  
  
The measurements are performed in vacuum at room temperature.  Uniform RF magnetic fields along the $y$- and $z$- directions are applied using small copper wire coils wound on CNC-machined PEEK coil forms.  The coil field distributions in (A/m)/mA of coil current are calculated by finite-element RF modeling \cite{COMSOL}, and the experimental field strengths determined in separate measurements by placing a commercial in-line RF current probe (Tektronix CT-6) close to an identical coil structure outside the vacuum chamber.  The phases of the RF magnetic fields at the sample are determined one drive coil at a time with the current probe.  A small home-made inductive sense coil (2 turns, $\sim 1.5\thinspace$mm diameter) is also used to measure the relative field amplitudes from the two coils near the sample position, to confirm the estimate from the finite element calculations and current amplitude measurements.  The right-handedness of the field geometry is determined with the Hall probe using DC currents through the RF coils (left-handed field coordinates would introduce a minus sign for one of the torques). 

Uniform DC bias fields for the precision low-field torque measurements are applied using home-made hoop electromagnets.  Stronger DC fields as required for full hysteresis loops taking the YIG disks into saturation are applied using a NdFeB permanent magnet mounted on a lead screw driven rail, with the position and orientation of the magnet computer-controlled through a pair of stepper motors.  DC field strengths are continuously monitored with a 3-axis Hall probe during the measurements.  

Mechanical displacements driven by magnetic torque are recorded through interferometric modulation of reflected optical intensity from a 633 nm beam focussed near a silicon paddle edge.  Torque sensitivity is calibrated via thermomechanical displacement noise at the mechanical resonance, in the absence of RF torque drive.  The torsional nature of the mechanical modes used for measurement are characterized by finite-element mechanical modeling \cite{COMSOL}, and confirmed through spatial maps of the signal generated by raster scanning the sample position under the laser focus.  The moment of inertia of the device is asymmetric about the torsion axis, on account of the mounting of the YIG disks on top of the paddles.  This asymmetry causes some hybridization of torsion and flexing motions, through angular acceleration and centripetal force.  The effect on torque sensitivity is a small percentage reduction, as estimated using a three-spring toy model of the coupled modes (see supplementary material section S5 \cite{supplement}).    

A characteristic full hysteresis measurement acquired through cross-product torque is shown in Fig.~\ref{Fig2}a, for a double paddle sensor with a disk on each paddle.  The measurement is run with a phase-locked loop to maintain the drive frequency at the same point on the mechanical resonance curve as the bias field is swept, correcting for the stiffening of the torsion mode as the Zeeman energy minimum deepens at high bias field (the drive frequency range across the measurement is 2.7882 MHz to 2.7905 MHz).  The hysteresis measurement is used to constrain the magnetic aspect ratios of the disks, by comparison to micromagnetic simulation (Fig.~\ref{Fig2}b).  Thinner disks magnetize more easily (higher positive slope at low field\cite{Cowburn1999}), and exhibit a proportionally slower decrease of torque (smaller negative slope) at high field where the $m_x$ moment is saturated; the rate of decrease is governed by the $z$-direction susceptibility, $\chi_z$, which is larger for thicker disks.  The intersection point of linear fits to the low-field and high-field torques yields a characteristic field that depends sensitively on the $h/d$ aspect ratio, yielding $0.23 \pm 0.01\thinspace$ for this sample. In the hysteretic region between 12 and 18 kA/m where the spin texture transitions between vortex and quasi-uniform, the simulation does not represent the experiments accurately; much longer-running simulations incorporating thermal activation and the effect of magnetic edge roughness would have to be performed (see Section~\ref{SecVII}), and in addition the differences between the two individual disks accounted for.  Away from the hysteretic region, at low and high fields where in both cases the spin texture is well-defined, the simulations are mostly very well behaved.  The simulation in Fig.~\ref{Fig2}b includes a constant offset field $H_y = 2\thinspace$kA/m, to avoid the difficulty relaxing fully to equilibrium around $H_x = 23\thinspace$kA/m seen in Fig.~\ref{Fig7}a when $H_y$ is within 200 A/m of zero.

The peak measured and simulated torques are in the ratio $66\thinspace\textrm{aNm}/89\thinspace\textrm{aNm} = 0.74$.  This indicates that within the magnetically-dead layer caused by Ga beam damage during the milling (see supplementary material section S1 \cite{supplement}), the magnetic disks are smaller in linear dimension than the simulated disk by $\sqrt[3]{0.74} = 0.9$, or approximately $1.8\thinspace\mu$m diameter by $0.42\thinspace\mu$m thick.

\section{\label{SecIV}Low bias field vortex state Einstein-de Haas effect}

\begin{figure*}
\includegraphics[width=16.4cm,keepaspectratio]{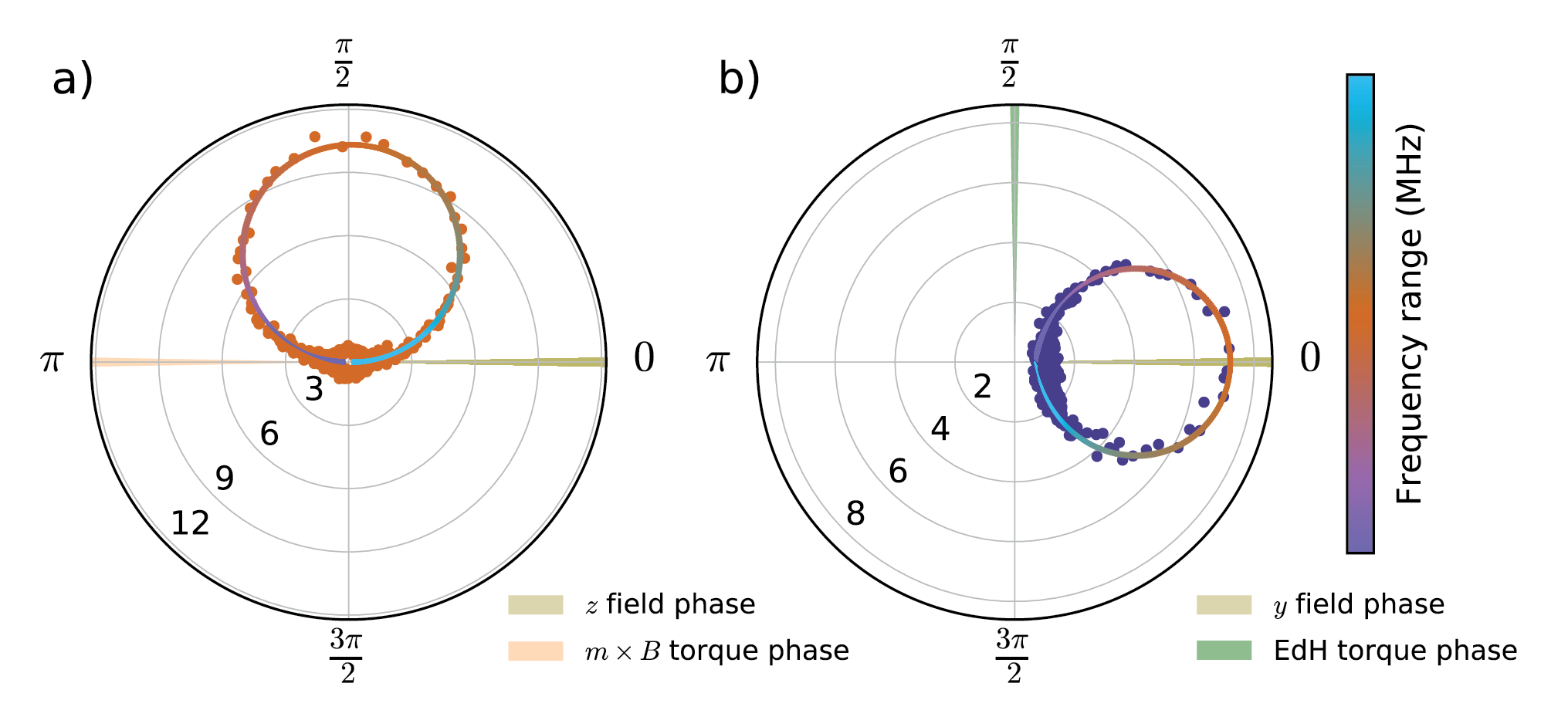}
\caption{\label{Fig3}Polar plots of the magnitude and phase of the signals versus frequency across the torsion resonance, for both RF drive field directions.  The solid lines are fits, and through their color encode the drive frequency (color bar on the right).   The drive field phases are at zero radians in both cases.  a) The expected cross-product torque phase is $\pi$ radians.  The  $H_z^\textrm{RF}$-driven signal starts in-phase with the torque below resonance, and develops a phase lag of $\frac{\pi}{2}$ at resonance.  b) The expected EdH torque phase is at $\frac{\pi}{2}$, and the mechanical signal driven across resonance by $H_y^\textrm{RF}$ behaves as in panel a) but with the overall phase rotation indicative of the EdH effect.  The swept frequency range is 2.75 to 2.84 MHz (simultaneous drives with $y$- and $z$-drive frequencies separated by 3.142 kHz), and the bias magnetic field $H_x = -200$ A/m.}
\end{figure*}

Comparative frequency sweeps through the mechanical resonance, driven by $H_z^\textrm{RF}$ (cross-product torque) and $H_y^\textrm{RF}$ (EdH) are shown in the polar plots of Fig.~\ref{Fig3}, at a low $H_x$  bias field where the two torques have similar magnitudes.  The mechanical signal traces out an accurate resonance circle in both cases.  The drive \textit{field} phases as determined by the current probe establish the 0 radian reference of the phase angle axis and are indicated by thin wedges.  The additional color markers show the corresponding cross-product and EdH drive \textit{torque} phases predicted by Equations 2 and 3.  Far below the resonance frequency, the torque signals begin in phase with the torque drive, as expected.  The $\pi/2$ rad phase lag of the response at the peak (maximum magnitude of the resonance circle) brings the EdH-driven response back into phase alignment with the drive field on resonance, highlighting the distinct origin of this torque.  There is a small constant offset on the measurement with the y-coil drive, arising from a radiative RF coupling between that coil and the photoreceiver.  The resonance circle illustrates that the crosstalk phasor is constant and can be separated from the mechanical signal. 

\begin{figure}[htb!] 
\includegraphics{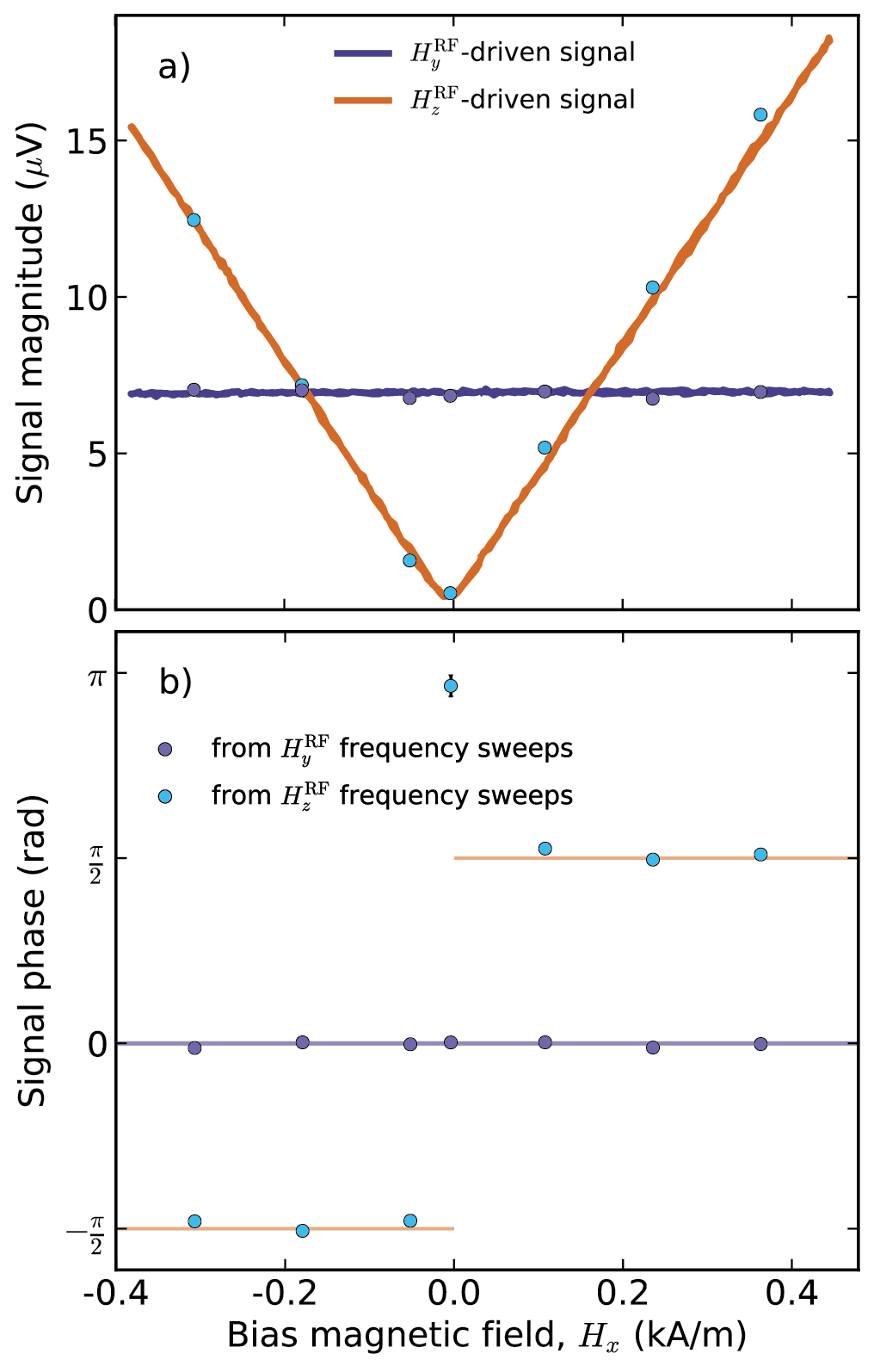}
\caption{\label{Fig4}Simultaneous cross-product and EdH torque measurements versus DC bias field, in the low field range from zero bias where the EdH torque dominates, and through the crossover of torque magnitudes for both polarities of bias field.  This illustrates a) the measured field dependence of the torque magnitudes, as predicted from Fig.~\ref{Fig1}b, with b) the phase information, from the lock-in measurements.  The drive frequencies are separated by 0.2 kHz for these measurements, and the 50 ms lock-in time constant ensures independent demodulation of the two signals.  The discrete points are confirmation measurements, from fits to frequency sweeps through the mechanical resonance acquired at fixed fields.}
\end{figure}

A signature measurement of the present study is shown in Fig.~\ref{Fig4}, which presents the amplitudes and phases of both radio frequency torques as a function of bias field on either side of $H_x = 0$ demonstrating that the EdH torque easily can exceed the cross-product torque over a significant bias field range, and also the expected phase relationships for both the cross-product and EdH torques (both measured at the peak of the mechanical resonance).  The linear-in-field dependence of the cross-product torque is observed, along with the bias field-independence of the EdH torque.  (The mechanically-resonant detection in this work does not permit a direct test of the latter's linear-in-frequency dependence.)  The discrete data points on the plot are determined from full frequency sweeps through the resonance at fixed bias fields, as in Fig.~\ref{Fig3}.  The phases determined this way are more accurate than those from the continuous field sweeps, which exhibit run-to-run variation of the phase of about 0.1 rad.  The phase-locked loop is not used here on account of the small signals. Measurements through full $360^{\circ}$ rotation of the in-plane field direction further highlight the qualitatively different behaviors of the two torques (see supplementary material section S3 \cite{supplement}).  

The results of Fig.~\ref{Fig4} can be used to estimate $g\prime$ for the YIG, through comparison with the simulation results of Fig.~\ref{Fig1}.  Whereas Fig.~\ref{Fig1} assumed equal amplitude $H_y^\textrm{RF}$ and $H_z^\textrm{RF}$ drives, in the measurement we have $H_y^\textrm{RF} / H_z^\textrm{RF} = 1.41 \pm 0.05$ (see supplementary material section S2 \cite{supplement}).  For comparison of the measurements to Eq.~\ref{eq:six}, we take one half of the bias field separation between the two points at which the EdH and cross-product torques are equal, and define that as the the measured crossover field, $H_x^\textrm{crossover}$ ($ = 170 \pm 5\thinspace$A/m), and therefore

\begin{equation}
1 = \frac{\mu_0 c_\tau \gamma\prime}{2\pi} \frac{H_x^\textrm{crossover}}{f}\frac{H_z^\textrm{RF}}{H_y^\textrm{RF}}.
\label{eq:seven}
\end{equation} 

\noindent The first quotient on the right-hand side, as based on the simulations of Fig.~\ref{Fig1} and the hysteresis measurement of Fig.~\ref{Fig2}, is determined to be (see Fig.~\ref{Fig1}d)

\begin{equation}
\frac{\mu_0 c_\tau \gamma\prime}{2\pi} = (1.3 \pm 0.1) \times 10^4 \frac{\textrm{m}}{\textrm{C}}\cdot g\prime.
\label{eq:eight}
\end{equation}

\noindent The experimental results, jointly with the simulations to account for magnetic anisotropy, consequently arrive at the determination 

\begin{equation}
\begin{split}
g\prime & = \frac{f}{H_x^\textrm{crossover}} \frac{H_y^\textrm{RF}}{H_z^\textrm{RF}} \frac{1}{(1.3 \pm 0.1)\times 10^4} \frac{\textrm{m}}{\textrm{C}}\\ & = 1.78 \pm 0.16.
\end{split}
\label{eq:nine}
\end{equation}

The $g\prime$ value above is 1.5 standard deviations away from the value close to 2.0 that is to be expected based on arguments that $g - 2 \approx 2 - g\prime$, where the spectroscopic g factor for YIG is very close to the free electron value (the original determination by Dillon from ferrimagnetic resonance measurements on YIG spheres was $g = 2.005 \pm 0.002$ \cite{Dillon1957}).  The dominant contributions to the uncertainty are the drive field strength ratio and the detailed magnetic shape and aspect ratio.  The advantages of the microscale specimen studied here are that it demagnetizes into the vortex state and enables direct comparisons with micromagnetic simulations.  It will be possible to obtain substantially improved accuracy in the determination of $g\prime$ while retaining the advantages of high mechanical frequency operation by using specimens on the order of $10\times$ larger. For these larger samples the magnetic dead layer introduces only a very small difference between the magnetic shape and the physical shape, and by using a split coil for $H_z^\textrm{RF}$ to reduce the dominant uncertainty in the drive field strength ratio.  

A possible source of systematic error in the present experiment is an in-plane shape anisotropy that the current measurements cannot characterize; this can be addressed in future work through the incorporation of a second axis of torque detection.  Note also that prior knowledge of the saturation magnetization, $M_s$, of YIG enters the analysis implicitly through the shape anisotropies from simulation.   When all the relevant anisotropies can be characterized experimentally, $g\prime$ determinations for individual specimens through combined EdH and cross-product torque measurements will be possible without any additional inputs.    

\begin{figure}[htb!] 
\includegraphics{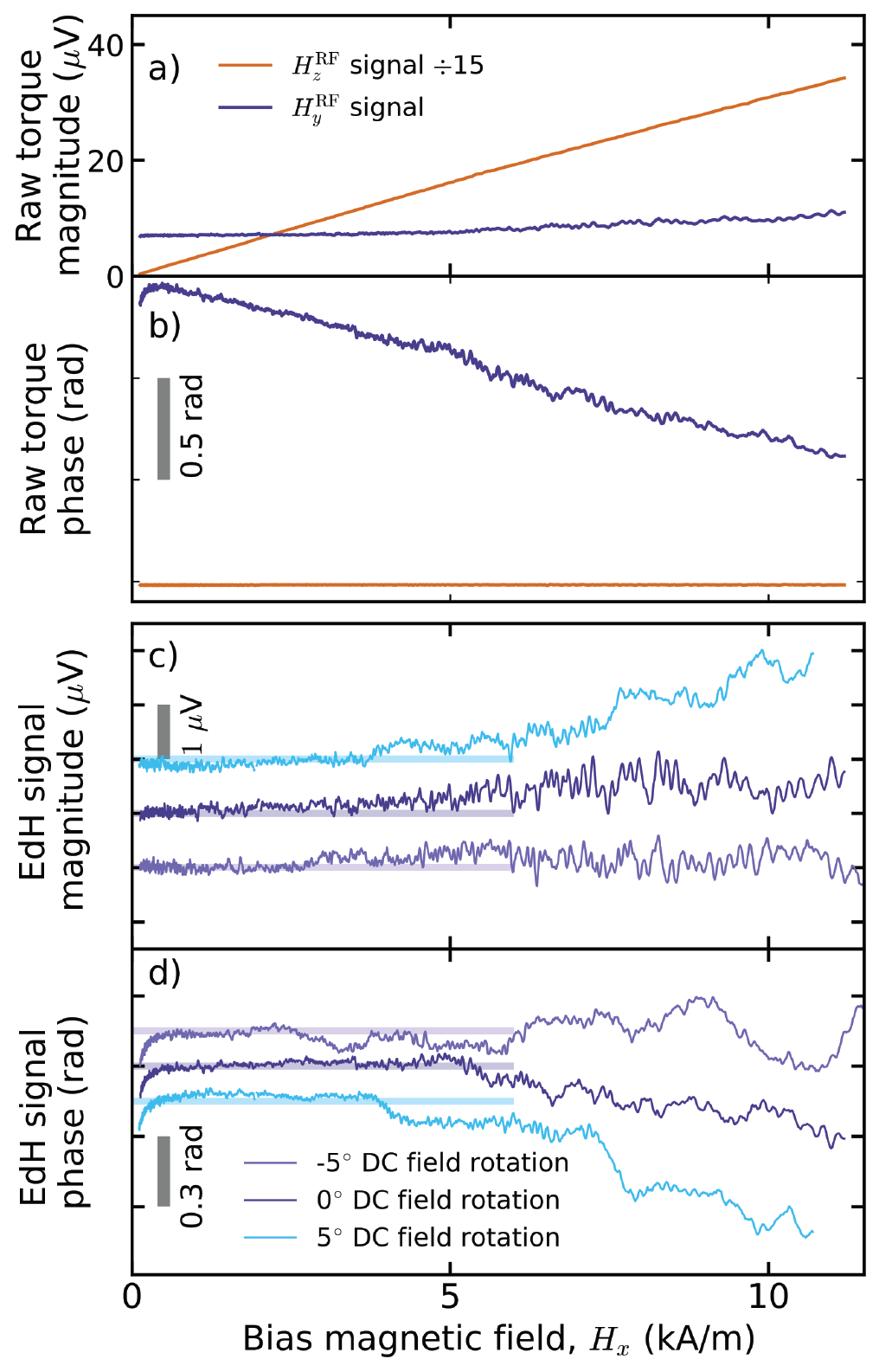}
\caption{\label{Fig5}Torque signals through most of the reversible, non-hysteretic bias field range where the disks remain in the vortex state, for different in-plane field angles (along $\hat{x}$, and at $\pm 5^{\circ}\thinspace$ from $\hat{x}$.  The cross-product torque (signal divided by 15) and raw $H_y^\textrm{RF}$-driven torque magnitudes and phases are shown in panels  a) and  b), respectively, for bias field along $\hat{x}$.  The EdH torque magnitudes and phases are shown in panels c) and d), with the different bias field direction traces offset for clarity (offsets of 1 $\mu$V and 0.15 rad in c) and d), respectively).  Indications of weak vortex core pinning are revealed by the field-angle dependence arising at bias field strengths above a few kA/m.  The signals are measured simultaneously, with the $H_z$-driven channel running in a phase-locked loop to account for the bias field-induced torsion resonance frequency shift.  The small downturn of $H_y$-driven phase near zero bias is from the PLL beginning to have difficulty holding lock on the smallest cross-product torque signals.}
\end{figure}

\section{\label{SecV}Higher bias field vortex state Einstein-de Haas effect}

Owing to the $\pi /2$ phase difference between the EdH and cross-product torques, it is possible to continue measurement of the EdH effect far beyond the crossover field, even with an RF field geometry in imperfect alignment with the mechanical coordinate system.  The machining and assembly inaccuracy of $\sim$10 mrad misalignment of the plane of the torsion resonators relative to the $y$-coil RF field direction at the sample is sufficiently small for easy separation of the unintentional admixture of cross-product torque from the raw data.  At this degree of misalignment, the cross-product torque has to develop to $100\times$ the level of the EdH torque before the unintentional admixture makes an equal magnitude contribution to the $H_y^\textrm{RF}$ coil-driven signal, a circumstance in which the two contributions still will separate easily and accurately in a quadrature phase-sensitive measurement.  

The experimental investigation of the EdH effect to much higher DC bias fields, while remaining in the vortex spin texture, is summarized in Fig.~\ref{Fig5}.  For these measurements the PLL is running on a slightly detuned signal driven by the $H_z^\textrm{RF}$ coil.  As the small admixture of cross-product torque driven by the $H_y^\textrm{RF}$ coil grows linearly with increasing $H_x$ bias field, and because this is adding in quadrature with the EdH torque, the \textit{phase} of the resultant raw signal varies linearly (Fig.~\ref{Fig5}b)  while its magnitude is largely unchanged (Fig.~\ref{Fig5}a; expected to vary quadratically as the admixture slowly grows).  The EdH-only signal is isolated from the raw signal by quadrature subtraction of the $H_z^\textrm{RF}$ coil-driven signal scaled to flatten the resulting phase at low fields.  A higher-order correction, from small misalignment of the $H_z^\textrm{RF}$ direction (versus the mechanical coordinate system) is negligible, owing to the extreme smallness of the unintentional EdH torque that the corresponding unintentional $y$-component drive gives rise to.  The small tailing-away of the extracted phase at very low bias field is an artifact from growing phase noise as the PLL approaches loss of lock.  

Unexpectedly, given the absence of features suggesting magnetic disorder (Barkhausen effects) in the cross-product torque, the idealized behavior of the EdH torque persists only to bias fields of $\sim 3\thinspace$ kA/m.  As seen in Fig.~\ref{Fig5} panels c and d, small departures from the low-field baseline of the EdH torque emerge over the field range 3 - 11 kA/m.  These variations depend sensitively on the in-plane bias field direction, strongly suggesting that they arise from magnetic disorder.  This indicates that there is a small amount of magnetic surface roughness in the as-fabricated sample, not large enough to become visible in the cross-product torque, but too large to remain invisible in the EdH torque.  

Why the distinction?  First, the highest energy density regions of the vortex core are at the surfaces, creating the likelihood of interactions with surface imperfections.  The core has significantly larger diameter in the center of the disk as compared to at the surfaces, owing to the thickness of the disk being approximately $25\times$ the dipole-exchange length in this case.  However, the $H_z^\textrm{RF}$ field driving the cross-product torque is a negligible perturbation on the spin texture; it induces a very slight `breathing' of the core diameter, but no modulation of the in-plane core position.  If the core is encountering pinning potentials in a weakly-disordered magnetic energy landscape, thermal fluctuations will drive hopping between neighboring energy minima and the RF cross-product torque will register the DC magnetic moment corresponding to the time-averaged core position.  Based on the characteristic energy corrugation even in more strongly pinning systems like polycrystalline permalloy at room temperature, the hopping rate is expected to be too fast in comparison to the measurement bandwidth (determined by the lock-in time constant and filter roll-off) for telegraph noise to be visible \cite{Burgess2013}.

In contrast, the $H_y^\textrm{RF}$ drive for the EdH torque induces a direct modulation of the core position at the drive frequency.  The angular frequency of the drive is 5 orders of magnitude higher than the largest measurement bandwidths of the present work, creating the possibility of thermally-activated hopping rates to come into range of the drive frequency as the core moves across the disk (the core equilibrium position being dictated by $H_x$).  In such circumstances a thermally-assisted, stochastic resonance-like \cite{Dykman1992, Gammaitoni1991} coherent motion of the core may result.  The consequent response to the $H_y^\textrm{RF}$ drive will exhibit enhanced amplitude and modified phase.  The emergence of these features only when the core is far enough from the center of the disk may be indicating that the disk fabrication somehow yields greater magnetic smoothness near the centers, or alternatively, that the magnetic disorder is primarily confined to magnetic edge roughness and interaction with the disk edges becomes more important at higher bias fields as the spin texture loses its circular symmetry.  The ion-milling fabrication procedure is more likely to cause an irregular disk perimeter and hence magnetic edge roughness in comparison to surface roughness.  Residual small scale surface roughness will remain after polishing the YIG wedge, but what this looks like magnetically, after the development of the dead layer from ion damage, can at the present time only be speculated upon based on the measured magnetic behavior. 

\section{\label{SecVI}High bias field EdH through hysteretic transitions\\ and in the quasi-uniform state}

\begin{figure}[htb!] 
\includegraphics{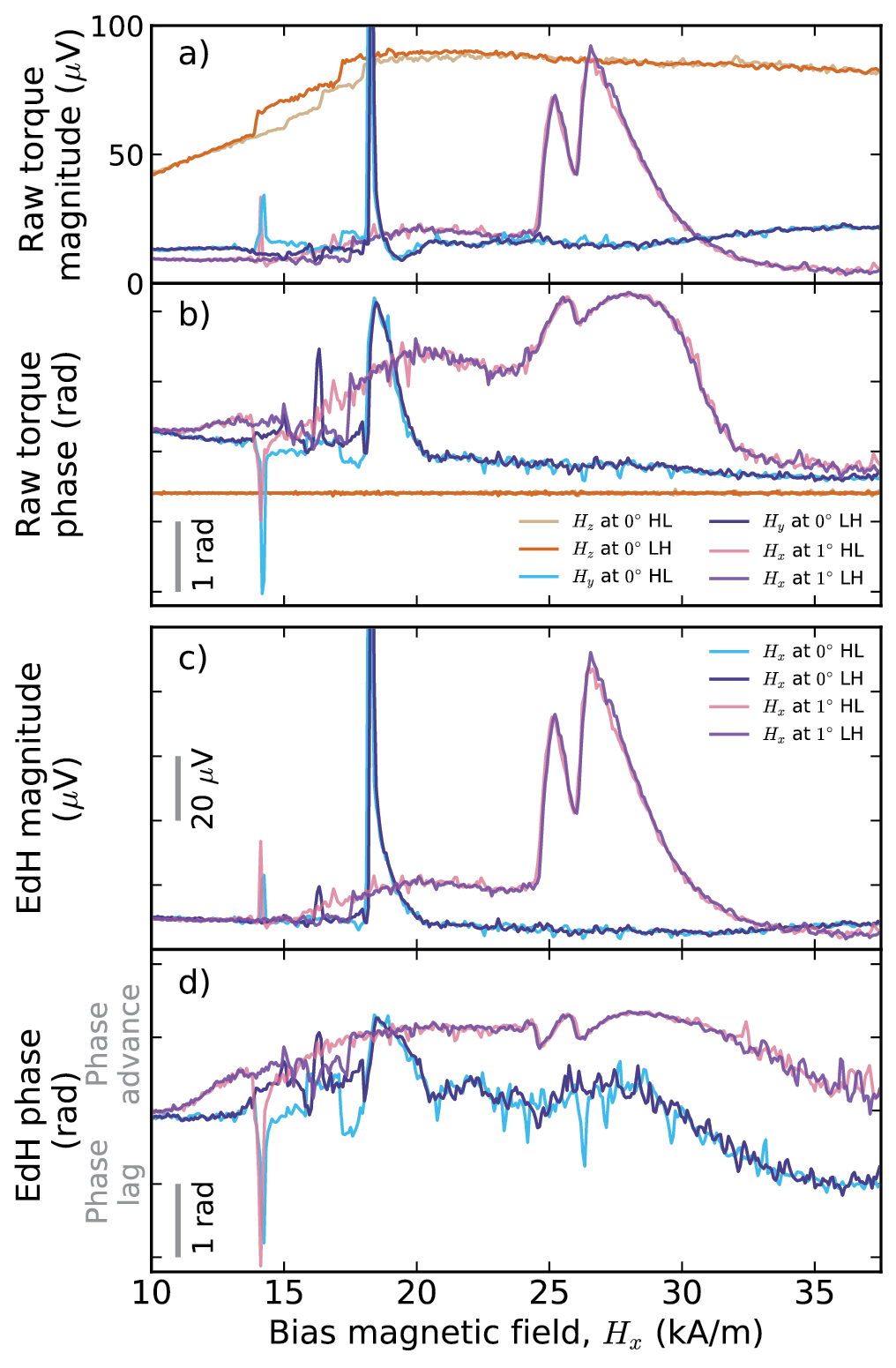}
\caption{\label{Fig6}The high bias field, hysteretic region of transitions in and out of the vortex state, as reflected in both EdH and cross-product torques, simultaneously measured. The direction of the bias field sweep is from low (L) to high (H) and back. The raw torque magnitudes and phases are shown in panels a) and b), for two in-plane bias field directions differing by only $1^{\circ}$.  The EdH torque magnitudes and phases, as determined by the procedure for removing the small cross-product admixture in the $H_y^\textrm{RF}$-driven signal, are shown in panels c) and d).  The features in the EdH torque depending strongly on bias field strength and direction are attributed to magnetic edge roughness, thermal activation, and possibly the beginnings of domain wall-like resonances. The clipped sections in panel a) and c) extend to 150 $\mu$V.}
\end{figure}

Figure~\ref{Fig6} shows the continuation of the EdH torque data into the quasi-uniform spin texture.  The unaltered magnitude and phase data from the $H_y^\textrm{RF}$ and $H_z^\textrm{RF}$ coils are shown (panels a and b) together with the resultant EdH signal after removing the cross-product admixture from the $H_y$ coil signal (panels c and d). At high fields, the EdH torque begins to decrease slowly, on account of the $y$-direction RF magnetic susceptibility, $\chi_y$, slowly decreasing as the saturated moment direction becomes more strongly anchored along $\hat{x}$ in larger bias fields.  As noted earlier for the case of $\chi_z$, determining pure susceptibilities via EdH torques creates the possibility of extracting saturation moments from measured cross-product torques in high bias fields.   

Surprisingly, in the hysteretic transition region between the vortex and well-saturated magnetization states, large peaks in the EdH torque (from peaks in the RF $y$-susceptibility) are observed over some ranges of $H_x$.  The same test for signatures of magnetic disorder applies here as in the vortex state, small changes of the in-plane direction of the DC bias field.  Instead of changing the trajectory of the vortex core and thereby having it explore a different cross-section of the disordered landscape, here a small rotation of where the most non-uniform regions of the spin texture are located around the disk perimeter is being effected.  This gives rise to the analogous mechanism of enhancement of the susceptibility by synchronization to the RF of thermally-activated hopping between pinning sites, and where now the pinning broadly characterized arises from magnetic edge roughness.  

The bias field rotation tests reveal a strong sensitivity to direction for the peaks beyond the last irreversible `annihilation' step in the cross-product torque at 18 kA/m, which we therefore characterize as extrinsic features.  Additionally, a rotation-independent peak is seen in the returning branch of the hysteresis, just before the last irreversible `nucleation' step at 14 kA/m in the sweep from high to low field.  This intrinsic feature indicates a softening of the spin texture just before nucleation, and could also be thermally assisted.  Spin texture softening has been observed before in magneto-optical susceptibility measurements arrays of permalloy disks \cite{Burgess2010} and in torque-mixing susceptometry of a YIG disk \cite{Losby2014}.  

The contrasting phase signatures of the putative extrinsic and intrinsic EdH torque enhancements are another important feature.  Ordinarily, one expects to find only additional phase lag (beyond that of the resonant mechanics) as a response of the system becoming unable to keep up with the pace of the RF magnetic field drive.  For example, when the drive frequency is more than negligible in comparison to the fundamental resonance of the system and the dissipation is significant, a measurable phase lag can develop simply from the physics of a damped resonator driven below resonance.  Torque-mixing magnetic resonance spectroscopy (TMRS) has been performed on a sibling specimen on the same chip with a similar, single YIG disk to confirm the presence of the expected gyrotropic vortex resonance (see supplementary material section S5 \cite{supplement}).   With the gyrotropic mode at 50 MHz and the RF drive therefore at $\sim 5\%\thinspace$ of the magnetic resonance frequency, an EdH torque phase shift of $\le 0.05\thinspace$ rad can develop, which is consistent with the low field observations of Figs.~\ref{Fig4} and~\ref{Fig5}.  The unpinned gyrotropic mode frequencies decrease as the DC bias field increases toward the vortex annihilation transition.  In combination with high damping, this will contribute to the growing phase lag observed for some bias field directions at higher fields.  The TMRS data show a fading-out of the fundamental gyrotropic mode at higher field, either an indication of stronger effective damping or, conceivably, of the onset of pinning making the gyrotropic frequencies increase \cite{Compton2006, Chen2012}.  The pinning potential in combination with thermal activation and the RF in-plane drive then gives rise to phase shifts, as already described.  

The TMRS data also reveal an even lower frequency spin resonance mode in the upper hysteresis branch (sweeping down from high field) over a narrow field range just above the vortex core nucleation transition.  The signals beyond 15 kA/m in Fig.~\ref{Fig6} make it clear, however, that the overall effect of magnetic disorder on the EdH torque phase is more complex.  In particular, it is observed that the signal phase can \textit{advance} relative to the drive phase as well as lag.  This unexpected phenomenology can be motivated by noting that it is possible, in a disordered 2D energy landscape, for the arrangement of neighbouring pinning sites to give rise to hops that reverse the sign of the differential (AC) magnetic susceptibility relative to the case in the absence of pinning.  This has been observed for the susceptibility component parallel to the bias field \cite{Burgess2013} and can also occur for the in-plane susceptibility perpendicular to the bias field.  In the case of a vortex texture, this requires a core trajectory exhibiting minor hysteresis around a local peak in the energy landscape, but where the local minimum occupied after a ``forward'' push on one side of the peak is farther back on the other side; and vice-versa -- effectively changing the sign of the differential susceptibility.  A related phenomenology could manifest for small closure domains.  

Additional insight will come from measurements of frequency and temperature dependencies of these phenomena in future experiments.  It will be highly informative also to augment the studies with an additional axis of torque detection \cite{Hajisalem2019}.  Bearing in mind that an in-plane cross-product torque must have an anisotropy as its foundation, a sensitive angular dependence similar to that found for the EdH torque would point directly to an extrinsic source such as magnetic edge roughness.  The corresponding effective torsion constants characterizing magnetic energy change versus bias field angle around such defects can have either sign depending upon whether the extremum is a local energy maximum or minimum.  Finally, it is possible that low frequency domain wall-like spin resonances could exist on a YIG disk periphery and couple to the mechanics, even at these comparatively small mechanical frequencies \cite{Saitoh2004}.  These would be accompanied by a net angular momentum absorption from the RF drive, which through suitable modulation could be detected via a torsion resonance along $\hat{x}$ (the direction of net longitudinal spin relaxation for such a mode in the present field geometry).  An interesting open question is whether there is net angular momentum absorption for the case of stochastic resonance.

\section{\label{SecVII}Simulations}

\begin{figure}[htb!] 
\includegraphics{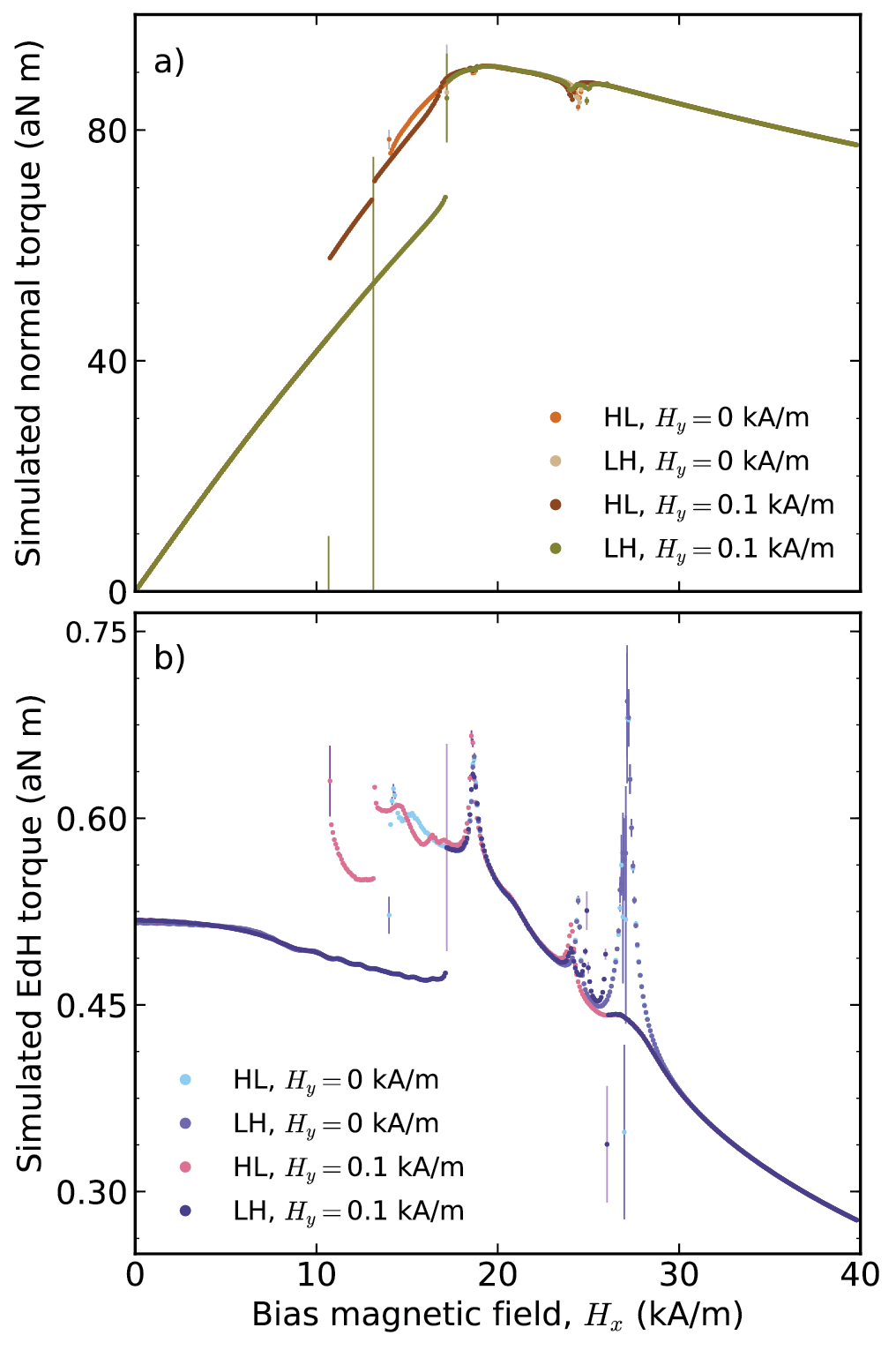}
\caption{\label{Fig7}Cross-product and EdH torques around the full hysteresis loop from micromagnetic simulation, using the same dimensions and drive field amplitudes as Fig.~\ref{Fig2}, for two bias field trajectories differing by a small $H_y^\textrm{DC}$ offset (see legend). The experimental ratio $f/g\prime = 2.79 \textrm{MHz}/1.8$ is used to calculate the EdH torque in b). HL, LH indicate the field sweep direction (high to low, low to high).}
\end{figure}

Figure~\ref{Fig7} summarizes the current status of combined micromagnetic simulations of EdH and cross-product torques.  For each bias field, $H_x$, in the micromagnetic simulations, small loops are computed for $H_y$ and (separately) $H_z$, varying from 0 up to + 8 A/m, down to - 8 A/m, and back to 0.  Efficient equilibration at each net field value is accomplished through a combination of setting a high value of the Gilbert damping constant ($\alpha = 0.5$), running a short interval (200 ps) with the Langevin temperature term \cite{Leliaert2017} set to 50 K to avoid pinning in metastable states that can be caused inadvertently by the finite element grid, and finally by letting the system relax with the Landau-Lifshitz precession term disabled and T = 0 (MuMax3 relax() function \cite{Vansteenkiste2014}).  ``Raw'' simulation outputs are post-processed to extract the torques.  The slope of the calculated $\vec{m}\times \mu_0 \vec{H}$ versus $H_z$, multiplied by the experimental $H_z^\textrm{RF}$ drive field amplitude, yields the simulated cross-product torque (Fig.~\ref{Fig7}a).  For the EdH effect, the simulated RF $y$-susceptibilities are converted to torque using Eq.~\ref{eq:three} (Fig.~\ref{Fig7}b).  This approach to the simulations is predicated on the operating hypothesis that the mechanical frequency is low enough to avoid the necessity of accounting for any spin dynamics.  Said hypothesis includes the assumption that spin-lattice relaxation rates, which are outside the physics described by the Landau-Lifshitz-Gilbert equation, are high enough to assume that the phase and amplitude of the resulting \textit{mechanical} torque (the experimental measurable) is identical to that of the magnetic torque.  

As mentioned in reference to Fig.~\ref{Fig2}b, simulations of the low RF frequency range magnetic torques are most straightforward for the low and high field bias ranges, where the equilibrium spin texture determination is not complicated by the presence of numerous, nearly-degenerate configurations.  Simulated EdH torques for the vortex state reproduce the bias field-independent initial EdH torque as found in the experiments, at low fields.  For the highest simulated bias fields, from 30 to 40 kA/m (well above the vortex annihilation at 18 kA/m in simulation), the spin texture is uniform enough that the simulated torques are robust against small changes in how the simulations are configured (as in the also non-hysteretic, and in other words single-valued, 0 - 10 kA/m range of the vortex state).  It is important to bear in mind, when looking at the high field range of Fig.~\ref{Fig7}b, that the $y$-susceptibility remains significant after the $x$-magnetization has saturated.  A corresponding simulated EdH $x$-torque (parallel to the bias field) would decrease rapidly towards zero above 18 kA/m.  Similarly for Fig.~\ref{Fig7}a, the non-zero $\chi_z$ underlies the decreasing cross-product torque at high fields.  $\chi_z$ is smaller than $\chi_y$ on account of the shape anisotropy, and proportionally decreases even more slowly with increasing bias field for the same reason. 

The micromagnetic simulations also offer first glimpses into the more complex phenomenology of the magnetic response observed in experiment both within and neighboring the field range of hysteresis between well-defined spin textures.  The step-up of EdH torque after vortex annihilation in the simulations, and exhibiting hysteresis on the field-sweep down, is seen qualitatively also in the experimental behavior over the same range for the $1^{\circ}\thinspace$field rotation data in Fig.~\ref{Fig6}c.  The dramatic peaking of experimental EdH torque over some field ranges in the quasi-uniform texture also is echoed in the simulations, and with Barkhausen effect-like fingerprints (sensitive to bias field magnitude and direction) that can be traced to edge roughness.  The finite-element simulation employs a grid of small rectangular prisms to approximate the cylindrical disk, which incorporates into the model a simple mimic of anisotropic magnetic edge roughness sensitive to small in-plane bias field direction changes.  Bias field direction changes are mimicked in the simulations presented through constant offsets of the $H_y^\textrm{DC}$ field.  Corresponding experimental measurements also have been made by setting $H_x^\textrm{DC}$ at a fixed value and sweeping $H_y^\textrm{DC}$ on either side of $H_y = 0$.  The susceptibility peaks found in these simulations may be somewhat enhanced when the simulation encounters many nearly degenerate configurations and has more difficulty relaxing to equilibrium.  The error bars in Fig.~\ref{Fig7} are computed directly from the slope uncertainties returned by the linear fits to $\vec{m} \mu_0 \vec{H}$ versus $H_z$ and $m_y$ versus $H_y$ in post-processing of simulation output.      
Thus far, no simulation feature has emerged as a compelling candidate relatable to the `intrinsic' EdH torque peak found in measurements just before vortex nucleation.  Future simulations must address explicitly the evolution of the torques in the time-domain, as required in order to model phase shifts arising in slow (thermally activated) and fast (precessional) spin dynamics.  Very straightforward in principle, this will require in the range of $10\times$ to $1000\times$ more GPU cycles per simulation; shortcuts such as turning off the LLG precession term apply only for the efforts to simulate equilibrium behaviors.  To assist with the challenges posed by lengthier simulations, pinning effects could be added to the single domain wall model used by Jaafar and Chudnovsky\cite{Jaafar2009} to analyze the Wallis, Kabos and Moreland experiment\cite{Wallis2006}.  Similarly, the analytical model of vortex core pinning\cite{Burgess2014} could be extended to include the description of EdH torques.

\section{\label{SecVIII}Discussion}

The results presented here underscore the combined necessity and utility of incorporating phase-sensitive detection with EdH effect measurements at radio frequencies.  A great advantage, stemming from phase orthogonality of EdH and cross-product torques when the system remains in magnetic equilibrium, is the ability to separate the two in a quadrature measurement.  This will prove helpful also in AC torque studies of larger (including macroscopic) magnetic specimens at lower mechanical frequencies.  In practice, considerable care is required in the measurements to guard against other sources of phase shift that could cause systematic error in the EdH torque phase.  Because the amplitude peak of a resonance corresponds to the point of maximum phase-versus-frequency slope, the effects of mechanical frequency shifts as caused by the Zeeman energy of the sample changing with DC bias field, and from temperature drift of the sensor, must be removed.  One means of stabilizing the measurements against phase shifts other than those arising directly in the EdH torque, as demonstrated here, is tracking a cross-product torque simultaneously in the same mechanical mode, using a phase-locked loop.

Scaling a given sensor geometry to smaller linear dimensions both increases the resonance frequencies and improves the absolute torque sensitivity.   Since the absolute sensitivities of small torque sensors are sufficient to overcome the cubic decrease of sample volume with linear down-scaling \cite{Losby2018}, small devices are particularly effective at discriminating EdH from other effects.  Applied field uniformity is easier to achieve over small sample volumes as well, improving the isolation from unintentional torques and gradient forces.  

The twisting motion of the sensor induces a back-action on the magnetization through the Barnett effect.  The magnitude of this back-action is negligible here (and indeed would be invisible to the measurements where the simultaneous torque drives have a slight frequency detuning relative to one another), but it is useful to develop a feel for the numbers, for future reference.  The scale of the shift in ground state energy from rotation is $\mu_0 m H_\Omega$, where angular velocity $\Omega$ gives rise to the effective field (or, in Barnett’s words, `intrinsic magnetic intensity of rotation’) \cite{Barnett1915,Ogata2017,Arabgol2019}

\begin{equation}
\begin{split}
H_\Omega & = \frac{2 m_\textrm{e}}{\mu_0 e g\prime} \Omega \\
& = \frac{4\pi m_\textrm{e}}{\mu_0 e} \frac{f^\textrm{rot}}{g\prime}.
\end{split}
\label{eq:ten}
\end{equation}

\noindent The numerical prefactor is the inverse of the scale-setting factor from Fig.~\ref{Fig1}d, here $1/17588 = 5.686 \times 10^{-5}$ C/m.  For harmonic angular motion of the torque sensor this is a sinusoidal effective field with $f^\textrm{rot}$ representing the instantaneous rotational velocity in cycles per second.  In the present experiments, the angular displacement amplitudes remain less than 1 mrad even when driven by the maximum cross-product torques (see supplementary material section S5 \cite{supplement}).  For the resonant frequency of 2.8 MHz this corresponds to peak rotational speeds $f^\textrm{rot} \sim 2.8 \times10^3\thinspace \textrm{s}^-1$, yielding a characteristic $H_\Omega \sim 0.08$ A/m.  The scale of $H_\Omega$ is $10^{-3}$ times smaller than the RF field driving the mechanical motion.  This is not significant here, but must be borne in mind for experiments where, as an example, higher mechanical $Q$ yields larger displacement per unit driving field.   

It will be highly desirable in some future experiments to have a \textit{second}, orthogonal torque detection axis.  As already noted, these include the characterization of additional anisotropies, and the determination of when there is a net angular momentum absorption from the driving RF field.  Elaborating upon the latter,  detection of spin resonances (which may be viewed as frequency-specific anisotropies) via the EdH effect is a powerful and underutilized technique.  The results presented here demonstrate the possibility of direct detection of spin resonances through EdH torques along the axis of the RF magnetic field, through the magnetic susceptibility enhancement on resonance (and assuming that the sensor has sufficient torque sensitivity at that frequency).  In addition, in a conventional magnetic resonance there is a steady flow of angular momentum into the spin system from RF absorption under continuous driving, effectively corresponding to a DC EdH torque parallel to the bias magnetic field and \textit{perpendicular} to the RF.  Modulation of this DC torque at some low mechanical frequency underpins the successful torque detection of resonance pioneered by Alzetta, Ascoli and Gozzini \cite{Alzetta1967, Ascoli1996}.  Torque-mixing magnetic resonance spectroscopy \cite{Losby2015}, on the other hand, is effectively an RF-modulated implementation of the direct RF-EdH measurements reported here.  The development of TMRS preceded the exploration of EdH manifestations at lower RF frequencies reported here, spanning frequencies from well-below to those approaching the lowest magnetic resonance mode.  Here, the mechanical (and hence RF drive) frequency remains constant while the frequency of the lowest magnetic resonance is tuned via the DC bias field (and in a manner dependent upon the spin texture).  Implementing the measurements with a second torque axis will enable direct intercomparisons between the different detection modalities, with the possibility of shedding new light on spin dynamics including spin-lattice relaxation mechanisms and their anisotropies.  It will also become possible to measure $g$ and $g\prime$ in a single mechanically-based experiment on the same sample, enabling important questions from the earliest days of spin dynamics \cite{Kittel1949,vanVleck1951} to be revisited for materials of contemporary interest.

\section{\label{SecIX}Summary}

The Einstein-de Haas experiment was a milestone in magnetism, demonstrating for the first time the anticipated intrinsic connection between net magnetization and mechanical angular momentum.  The expected angular momentum change upon reversing magnetization within the Amperian current model for magnetic moment was small, and drove Einstein and de Haas to undertake a mechanically resonant AC measurement wherein the angular displacement of a torsion balance would grow upon synchronous alternation of the poling direction of a supported magnet.  With the torque felt by the sensor proportional to the time rate of change of angular momentum, the challenge of discriminating the EdH torque from unintentional magnetic torques arising through unbalanced magnetic force gradients is exacerbated at low mechanical frequencies, possibly accounting for the original publication reporting a torque approximately $2\times$ larger than the true value determined in later experiments.  At the much higher frequencies of nanomechanical resonances, the EdH torques are much larger in proportion and it becomes easier, in a relative sense, to engineer the DC and RF applied field geometries required for adequate suppression of artifacts from other sources of mechanical drive.  All else being equal, the ratio of EdH to other magnetic torques increases linearly with the mechanical frequency.

As micro- and nanoscale torque sensors move to ever-higher mechanical frequencies, it will become essential to ensure that all of the EdH torques are accounted for, surprisingly, almost a reversal of the situation with regards to potential misidentification of signals in comparison to the original EdH experiments.  The linear scaling with frequency of EdH torque magnitudes cannot continue without limit, however, and it is precisely through the breakdown of such scaling that the methods will become most powerful and interesting.  Indeed, a recent theoretical study of the role of phonon spins in the EdH effect has discussed the conditions for decoupling of this contribution \cite{Ruckriegel2020}. Direct mechanically-based studies of spin-lattice dynamics in certain ordered magnetic systems may be on the not-too-distant horizon. 

Together, these features could elevate Einstein-de Haas measurements to the level of a mainstream tool in nanomagnetism, directly applicable to sensitive measurements of magnetic susceptibility in anisotropic systems, and ultimately even for the determination of spin-lattice relaxation times in magnetically-ordered (non-paramagnetic) states.  The radio frequency behaviors of EdH torque represent another example of the opportunities to expose new physics created by nanoscience, beyond miniaturization of earlier work.  Pure spin-mechanical, torque-mediated measurements, in a magnetism lab-on-a-chip implementation as envisioned by Moreland \cite{Moreland2012}, are poised to mine magnetic information in surprising depth.  

\begin{acknowledgements}
The authors gratefully acknowledge support from the Natural Sciences and Engineering Research Council (Canada), the Canada Foundation for Innovation, the Canada Research Chairs program, the National Research Council (Canada), and the University of Alberta.  We thank Dave Fortin for his rendering of the paddle device in Figure 1. We are also grateful to Katryna Fast and John Thibault for helpful conversations.   
\end{acknowledgements}

\bibliography{RF_EdH_refs-v5-30apr2020}

\end{document}


\title[Supplementary Material for The Einstein - de Haas effect at radio frequencies in yttrium iron garnet]{The Einstein - de Haas effect at radio frequencies in and near magnetic equilbrium:\\ Supplementary material}

\author{K. Mori$^1$}
\author{M.G. Dunsmore$^1$}%
\author{J.E. Losby$^{1,2}$}%
\author{D.M. Jenson$^1$}%
\author{M. Belov$^2$}%
\author{M.R. Freeman$^1$}

\email{mark.freeman@ualberta.ca}

\affiliation{%
$^1$ Department of Physics, University of Alberta, Edmonton, Alberta T6G 2E1, Canada
}%

\affiliation{%
$^2$ Nanotechnology Research Centre, National Research Council
of Canada, Edmonton, Alberta T6G 2M9, Canada
}%

{
\let\clearpage\relax
\maketitle
}

\renewcommand{\theequation}{S\arabic{equation}}
\renewcommand{\thefigure}{S\arabic{figure}}
\renewcommand{\thesection}{S\arabic{section}}

\section{YIG disk micromachining and mounting on torque sensors}

\begin{figure*}[htb!] 
	\centering
	\includegraphics{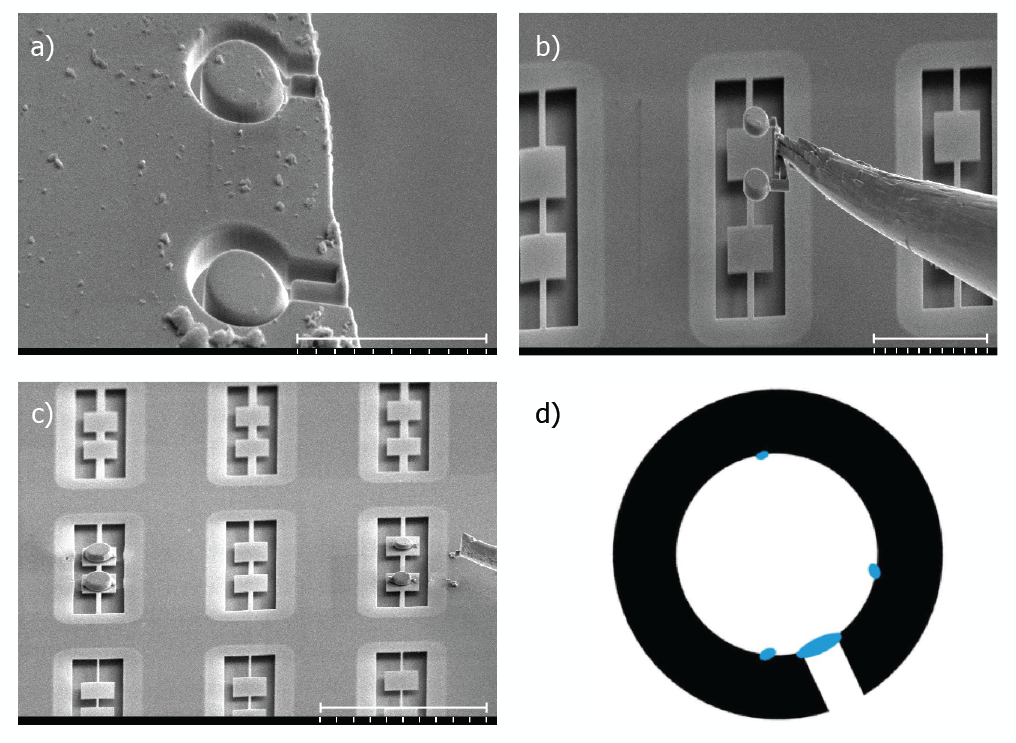}
	\caption{\label{fig:S1}%
	The focused ion beam YIG milling and nanomanipulation transfer process.  a) A pair of disks with handles, after outlining by FIB milling and still attached to the YIG plate. Scale bar = $5 \mu$m.  b) A disk pair attached to the nanomanipulation probe and poised for landing on a double-paddle torque sensor. Scale bar = $10 \mu$m.  c) A section of the pre-fabricated torque sensor array with two completed devices.  Scale bar = $20 \mu$m.   d) Geometry of the beam-induced welds fastening the YIG disks to the host torque sensor silicon paddles at three spots around each disk perimeter, and of the cut-off of the disk nanomanipulation handle (all indicated in blue).  The dark split-ring is the bitmap pattern used to mill the disk out of the thin YIG plate.}
\end{figure*}

Single-crystal YIG disks of diameter approximately 2.2 $\mu$m and thickness approximately 0.6 $\mu$m are milled out of a free-standing YIG plate approximately 0.8 $\mu$m thick.  These mesoscale disks are ensured to self-demagnetize to vortex ground states.  The YIG plate, from a liquid phase epitaxy-grown thick film with $\langle 111 \rangle$ crystallographic orientation \cite{Shin-Etsu}, was prepared by mechanical polishing of a small die sawn from a YIG/GGG/YIG wafer.  The micromachining process is summarized through the images in Fig.~\ref{fig:S1}.  A Hitachi NB5000 dual scanning focused ion/electron beam microscope system cuts disks from the thin YIG film with the Ga$^+$ beam (40 kV, 0.7 nA) and transfers in-situ onto prefabricated torsional resonators via nanomanipulation and beam-induced tack-welding (0.07 nA probe current with tungsten metallorganic precursor gas) \cite{Compton2012}.  The walls of discs are sloped as a consequence of the spatial profile of the Ga$^+$ flux being partially imprinted on the edge as the beam cuts through the YIG. The 0.07 nA probe cuts off the YIG handle used to temporarily connect the disks to the nanomanipulation probe.  The principal shortcoming of this method of fabrication is damage near the surface, manifesting as a magnetically-dead layer with characteristic thickness in the range of 50 nm \cite{Fraser2010}.  The microscopic mechanisms for inactivating the YIG magnetism are thought to include lattice damage from passage of the Ga beam and from local heating, and compositional change from Ga implantation.  A full microstructural characterization has not been performed.

Both the welding and the handle cutting are potential sources of magnetic edge roughness introduced through the sample preparation procedure (Fig.~\ref{fig:S1}d).  Smaller limitations to magnetic edge smoothness will arise from fluctuations in beam current and beam pointing, during the milling procedure.  Some top and bottom magnetic surface roughness is also to be expected, on the top from the pixel dwell time and consequent ion damage from the probe scan , and on the bottom from residual roughness and possibly tiny YIG particulates from the polishing process and/or redeposition from YIG during FIB milling.  

In the future, to establish a magnetic geometry with much less fractional uncertainty arising from the magnetic damage caused by the Ga$^+$ beam,  significantly larger YIG disks will have to be employed.  The downside of larger disks is the increased difficulty they have self-demagnetizing into a well-defined magnetic ground state.  

\section{RF coil geometries, RF field magnitude and phase calibrations}

The primary criterion in design of the RF coil assembly used in this experiment is ensuring accurate parallelism of the sample plane and the RF field used to drive the Einstein-de Haas torques.  The admixture of cross-product torque accompanying the EdH signals is thereby kept to the small scale where it can be separated through data analysis (and in the low bias field range, neglected entirely).  The bias field range over which the EdH torques exceed the maximum cross-product torques is small enough here that the analogous EdH torque admixture from the cross-product torque RF driving field is always negligible; however, it should be noted that this will not always remain the case at higher RF frequencies.  

The quasi-Helmholtz $H_y^\textrm{RF}$ drive coil geometry described below also effectively eliminates several possibilities for mechanical signal artifacts that might arise from more localized RF field actuation such as would be less forgiving to inaccuracies in positioning of the sample relative to a `sweet spot' of the RF field, such as gradient forces from inhomogeneity of the RF field, unintentional electrostatic forces if there is trapped charge on the mechanical sensor, and back-action electromotive force leading to phase shifts of the drive field.  

\subsection{Coil design}

The two drive coils for this measurement are mutually perpendicular, owing to the orthogonality of the two torque mechanisms and the mechanical detection that here uses a single torque axis.  A Helmholtz-type split coil is used to generate the field in the in-plane ($y$) direction while a current loop is used to drive the field in the $z$ direction.   Design parameters are based upon a COMSOL finite element simulation, and the assembly constructed by milling a small block of polyether ether ketone (PEEK) to make a combination coil form and device chip holder.  PEEK is compatible with high vacuum, is easily machined and rigid enough to accurately hold the finished shape.  The RF drive coils are wound using 30 gauge magnet wire (300 $\mu$m diameter including insulation). The $H_z^\textrm{RF}$ coil consists of two vertically close-stacked loops each of 3.2 mm diameter, with the design position of the sample plane being approximately 0.8 mm above the center of the upper loop.  The $H_y^\textrm{RF}$ coil assembly comprises twelve rectangular turns in total, six on either side of the sample, with each turns approximately 4.7 mm wide and 4.9 mm high.  Each group of six turns begins 3.0 mm from the sample and ends 5.0 mm from the sample.  A second copy of the coil assembly is used to perform reference phase and amplitude measurements outside the vacuum chamber while the experimental measurements are being conducted, and allows the current probe to be placed close to the coils for minimal phase lag.    Confirmation magnitude and phase calibrations are performed on the vacuum chamber coil assembly after the sample is removed.

\begin{figure*}[htb!] 
	\centering
	\includegraphics{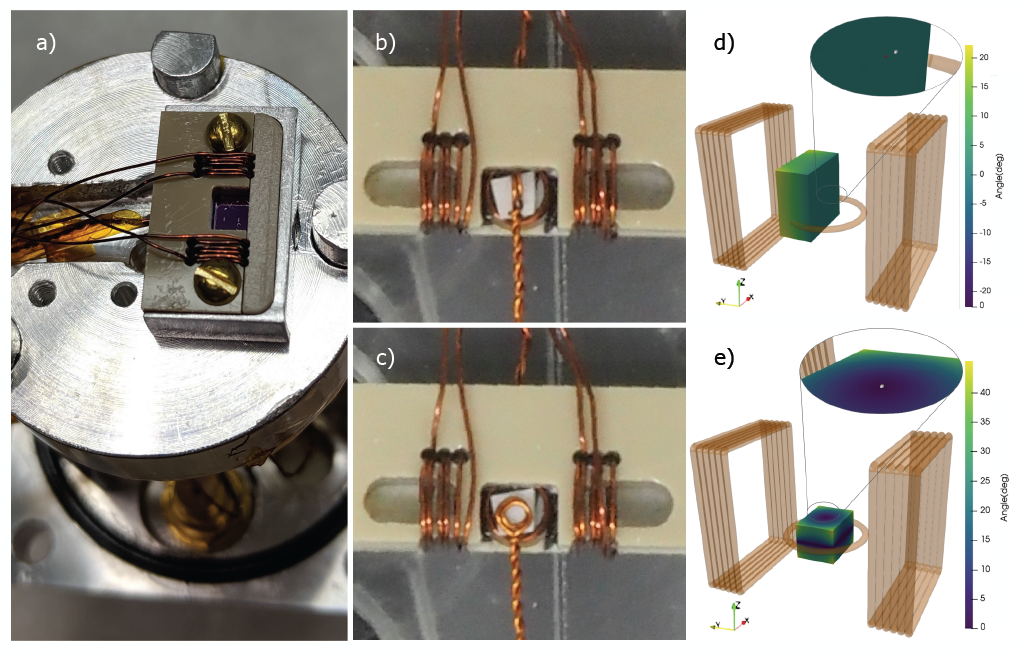}
	\caption{\label{fig:S2}%
	a) The RF coil assembly on its vacuum chamber stage, with the sample chip installed.  b) and c) The inductive probe in position for measurements of $H_y^\textrm{RF}$ and $H_z^\textrm{RF}$, for determining the ratio of RF driving field strengths. d) and e) COMSOL modeling of the RF magnetic field directions in the neighborhood of the sample position, for the $y$- and $z$- drive coils.}
\end{figure*}

\subsection{RF Magnetic Field Magnitude Ratio}

An essential input to the calculation of the magnetomechanical ratio, $g\prime$, from equilibrium Einstein-de Haas and cross-product torques is the ratio of the in-plane and out-of-plane RF drive field strengths.  The field strength at the sample position from each coil assembly is computed via Amp\`{e}re's law using the COMSOL AC/DC module \cite{COMSOL}.  The calculated RF field strength ratio is confirmed through inductive measurements with an RF magnetic field probe consisting of two 1.5 mm diameter turns of 30 gauge copper magnet wire with twisted pair leads.  A three-axis translation is used to position the inductive probe in the reference coil assembly, and can be rotated $90^{\circ}$ in-situ to switch between amplitude measurements for the respective coils (Fig.~\ref{fig:S2}b) and c)).  The electromotive force from the pickup coil is recorded with an SR844 RF lock-in amplifier.  The calculation is more accurate for determining the field ratio at the sample position, on account of the very small sample size and the vertical gradient in the $H_z$-field.  

The currents used in the experiments are measured using the Tektronix CT-6 current probe, and found to have a ratio $I_y^\textrm{RF}/I_z^\textrm{RF}$ = 1.12, differing from unity on account of slightly different RF power amplifier gains in the two branches.  Combining the measured current ratio with the field strength calculation yields 
\begin{equation}
\frac{H^{\textrm{RF}}_{y}}{H^{\textrm{RF}}_{z}}=1.41\pm 0.05.
\end{equation}

\noindent The COMSOL calculation also captures the small variation of RF field directions through a small volume in the vicinity of the sample (shown in Fig.~\ref{fig:S2} d) and e)) and confirms that uncertainty in RF field direction arising from inaccuracy in sample placement is negligible in this design. The only significant departure from ideality in alignment of the magnetic field and mechanical coordinate systems is the $\sim 10\thinspace$ mrad angle between the $H_y^\textrm{RF}$ and the sample plane from the installation of the chip in the holder, as found through the torque measurements and discussed in the main text.

In the future, for drive frequencies beyond a few tens of MHz, the methods described above for establishing drive field amplitude and phase at the position of the sample will be too inaccurate.  A much better approach within that scenario will be switching to in-situ magneto-optical current probing\cite{Freeman1997}, effective to tens of GHz and therefore presenting no bandwidth limitations within the range of mechanical frequencies that can be envisioned at present. 

\section{EdH and cross-product torque comparison through full rotation of the in-plane DC applied magnetic field direction (low field)}

\begin{figure*}[htb!] 
	\centering
	\includegraphics{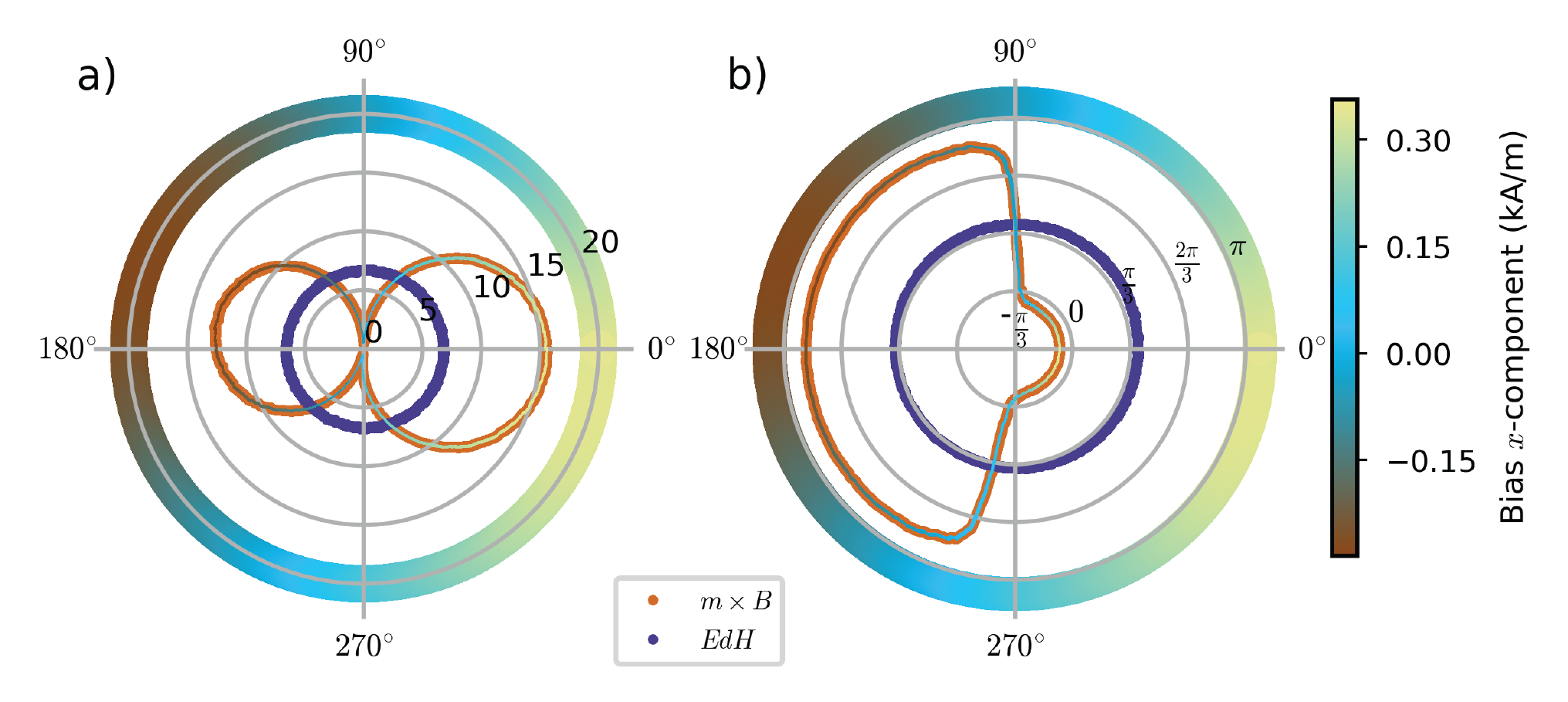}
	\caption{\label{fig:S3}%
	Low-field magnet rotation data (magnitude on the left and phase on the right), acquired with both torques measured simultaneously with drive frequencies detuned from each other by 200 Hz.  The two-inch cube NdFeB permanent magnet exhibits an asymmetry leading to different maximum DC field strengths in the two half-cycles, as shown by the color bar at the right (maximum field strength = $0.3578$ $kA/m$, minimum field strength = $-0.2844$ $kA/m$). The units of signal amplitude in a) are $\mu$V, and for signal phase in b) are radians.  The angle scale in degrees around the perimeter corresponds to the in-plane bias field direction.}
\end{figure*}

The contrasting phenomenologies of Einstein-de Haas and cross-product torques can also be highlighted through simultaneous measurements while rotating the DC bias magnetic field direction.  This is summarized in Fig.~\ref{fig:S3}, with signal magnitudes plotted in panel a) and signal phases in panel b).  The cross-product torque has two nulls in a full $360^{\circ}$ rotation, with the sign of the torque changing at each zero crossing, while the EdH torque magnitude and phase remain constant (in the raw $H_y^\textrm{RF}$-coil driven data, to within the level dictated by the small admixture of cross-product torque).  The in-plane field angle is encoded in the color scale around the perimeters of the polar plots, where the field direction is controlled for the measurements by rotating a NdFeB permanent magnet.  The maximum strengths of the bias field in opposing directions are unequal owing to asymmetry of the permanent magnet.  The strength of the $x$-component of the bias magnetic field is shown in the color bar at the right, which uses the same colormap as for the in-plane field angle.

\begin{figure*}[htb!] 
	\centering
	\includegraphics{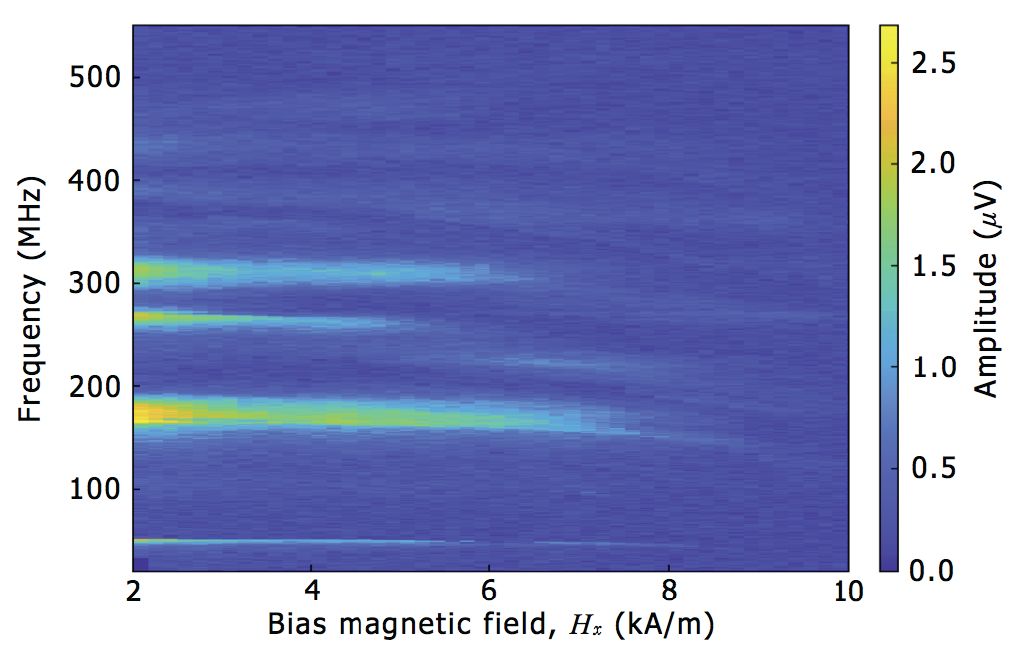}
	\caption{\label{fig:S4}%
	Mechanically-detected spin resonance spectra of a single YIG disk sample in the vortex regime.  The color bar shows the torque-mixing signal intensity.}
\end{figure*}

\section{Torque-mixing magnetic resonance spectroscopic characterization of spin dynamics in the disks}

To determine the frequencies of the lowest spin resonances of the micrometer-scale YIG disk, mechanically-detected spectroscopy measurements are performed with the torque-mixing method \cite{Losby2015}.  Of particular interest in the present context are the lowest frequency modes that are observed.  The fundamental gyrotropic mode in the vortex state is found at 50 MHz (Fig.~\ref{fig:S4}).  In the field sweep-down branch of the hysteresis (not shown), the lowest mode tunes down to below 40 MHz just above the spin texture transition back into the vortex state.

\section{Nanomechanical geometries and thermomechanical torque sensitivity calibration}

An ideal device for this kind of study would have its fundamental torsion resonance as the lowest frequency mechanical eigenmode, and with reasonable frequency separation from the fundamental flexural mode \cite{Losby2018}.  This condition is difficult to achieve in practice.    The construction of a nanomechanical torque device always results in some degree of mechanical asymmetry such that torsional displacement involves a component of circular motion of the center of mass about the axis of rotation.  At torsion resonances, the asymmetry introduces periodic acceleration that only can be countered by elastic flexion.  The resonance eigenmodes, therefore, become admixtures of twisting and translational motion.  

For the fundamental modes, the flexing displacement is in the same phase across the whole device, whereas the torsional displacement is in opposite directions on either side of the torsion axis.  As a consequence, the two displacements add together on one side, and subtract on the other.  The resulting motion of the paddle appears as if the center of rotation has shifted to one side.  

For optimal torque detection sensitivity, the motion should be monitored on the side where the two contributions to displacement add.  The interferometer can be calibrated by using the Brownian (thermomechanical) motion to determine the conversion factor from $\mu$V (output of photoreceiver from an optical intensity change) to pm (displacement at the device).  A generalized method of performing the calibration involves a finite element mechanical simulation to determine the total energy of the eigenmode from the displacement profile at a turning point of the motion, and equating this to $\frac{1}{2}k_\textrm{B}T$ \cite{Hauer2013}.  Equivalently, the kinetic energy at an instant of zero displacement can be used.  

To complete a calibration of torque sensitivity, the conversion factor from torque drive (here in aNm, the natural unit for this study) to pm.  To simplify this final calibration step for hybridized modes, a three spring analytical toy model is used to complement the finite element simulation.  The analytical model aids in estimating the strength of coupling between the torsional and translational motions, and then in determining the amplitudes of the two contributions to displacement arising from a pure torque drive.  

\subsection{Finite-element mechanical simulation and analytical toy model of hybridized twisting and flexing resonances}

\begin{figure}[htb!]
	\centering
	\includegraphics{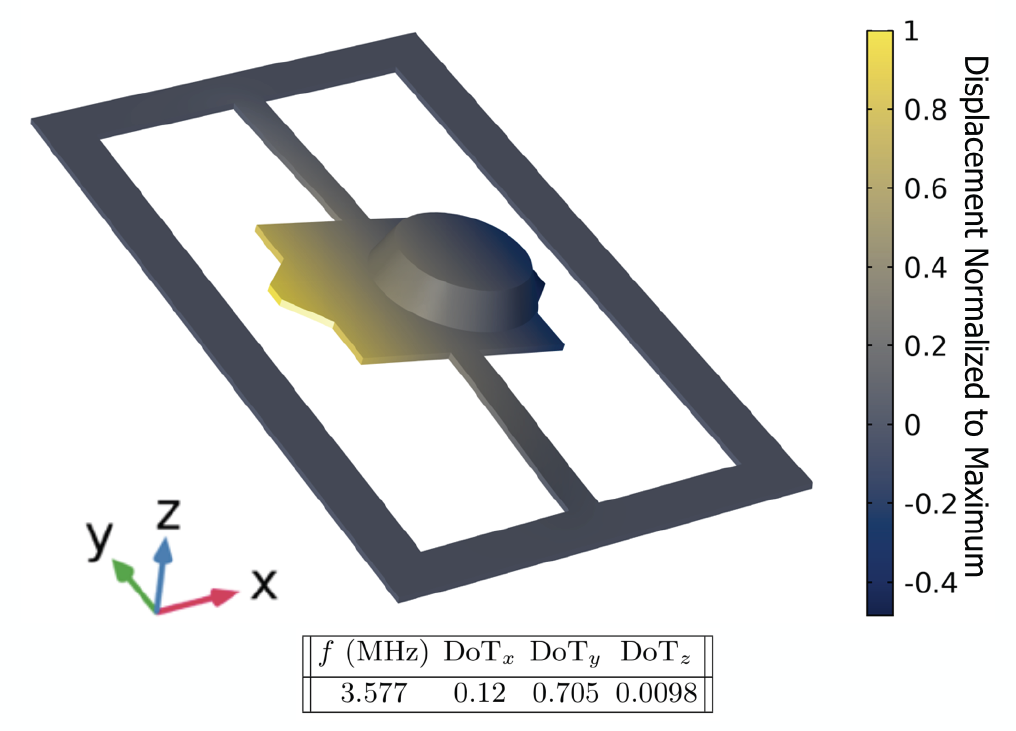}
	\caption{\label{fig:S5}%
	Simulation (COMSOL) of a torsional eigenmode with an easily visible admixture of flexing motion arising from significant asymmetry in the distribution of mass around the torsion axis. The color map represents normalized displacement.  The degree of torsionality for each axis corresponding to this mode is found in the table presented.}
\end{figure}

The admixture of torsion and translation is easily visible in Fig.~\ref{fig:S5}, which shows the lowest frequency twisting eigenmode from a COMSOL finite-element simulation for a single-paddle sensor loaded with a YIG disk.  The torsion arms have a uniform color map along their cross section, revealing the translational displacement occurring in concert with the clear rotational motion.  This introduces a perturbation on the torque sensitivity calibration.  

To characterize the efficacy of a given mode for torque sensing about a given axis, a Degree of Torsionality (DoT, $0 < \textrm{DoT} < 1$) figure of merit is defined.  A harmonic motion that exhibits admixture behaviour can be separated into energetic components relating to the rotational and translational motions.  Our definition of DoT takes the ratio maximum rotational kinetic energy to the total energy of the motion.  A pure torsion mode then has $\textrm{DoT}=1$ while a purely flexing mode has $\textrm{DoT}=0$.   Determining DoT from a COMSOL simulation involves obtaining the instantaneous rotational and transverse velocities and performing a volume integration over the entire resonator geometry, $\iiint_{\textrm{V}}\rho v^{2} dV$ (where $v$ separates into components of rotational and translational velocity, $v_{\textrm{R}}$ and $v_{\textrm{T}}$ respectively).  The total energy expression can be separated into three contributions,
\begin{equation}
E_{\textrm{Total}}=\iiint_{\textrm{V}}\rho v_{\textrm{R}}^{2} dV+\iiint_{\textrm{V}}\rho v_{\textrm{T}}^{2} dV+\iiint_{\textrm{V}}\rho v_{\textrm{R}}v_{\textrm{T}} dV.
\end{equation}
\noindent The cross term of rotational and translational velocity is generally smaller than the individual rotational and translational components.

\begin{figure}[ht]
	\centering
	\includegraphics{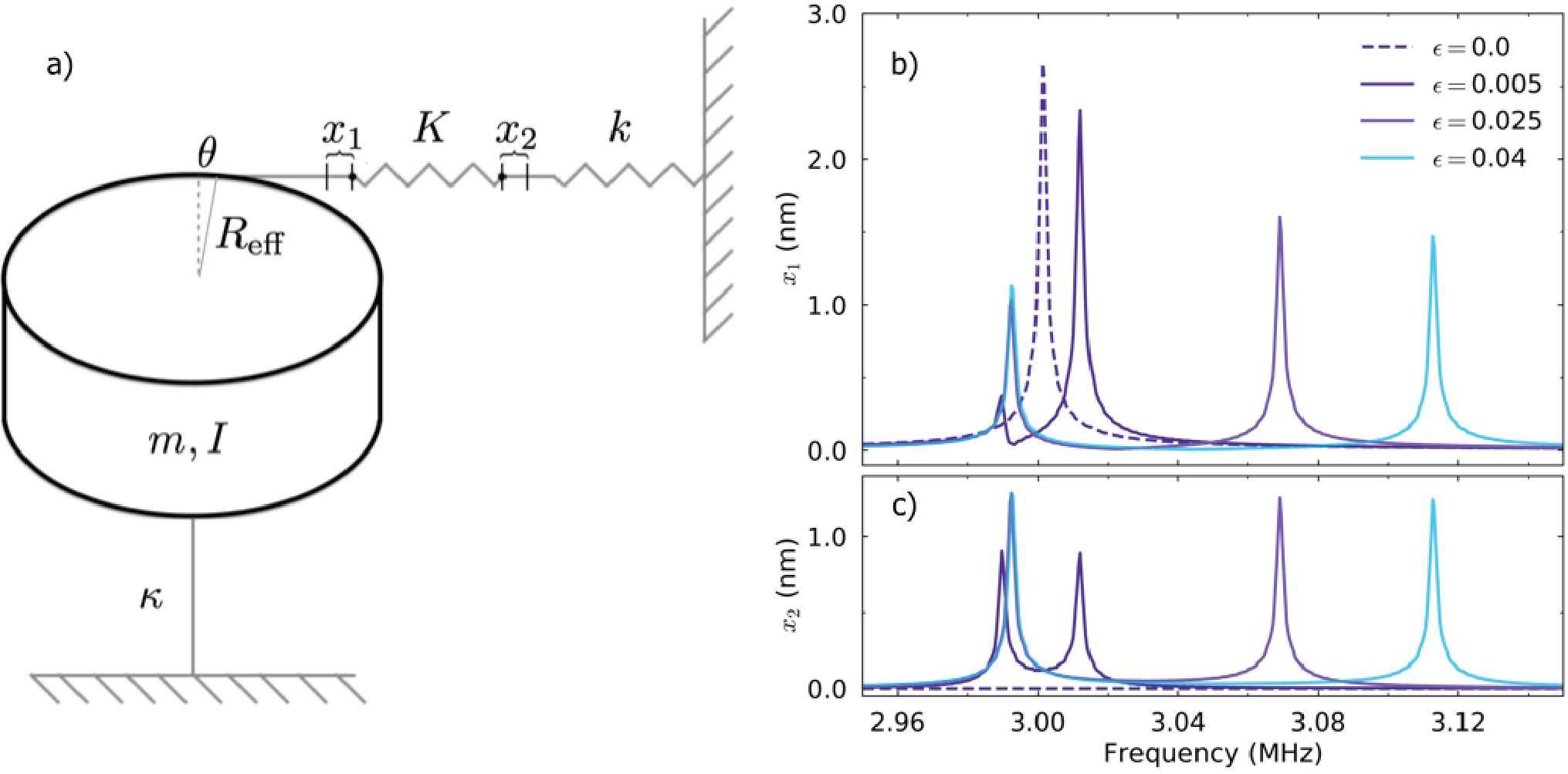}
	\caption{\label{fig:S6}%
	a) A torsion spring (illustrated as a straight line) with effective torsional spring constant, $\kappa$, coupled to a linear spring with spring constant, $k$, through a linear spring with spring constant, $K$. The respective mechanical frictions are $\gamma_{T}$ (torsional) and $\gamma_{L}$ (linear). Panels b) and c) show amplitude response due to a simulated torsional drive frequency sweep of the toy model spring system for various coupling factors. The solutions are obtained with a Python differential equation solver.  The dashed lines show the uncoupled case, where only the torsion spring responds to the drive.}
\end{figure}

The presence of mass asymmetry about the moment of inertia and close proximity of other deformational eigenmodes in frequency space is manifest in an admixture of mode shape. This admixture can be described as simultaneous torsional and translational deformational behavior. The response of such a system to a purely torsional drive may be explored by use of a toy model in which a torsional resonance dictated by a torsion spring of constant, $\kappa$, and moment of inertia, $I$, is coupled to a linear mass-spring system (spring constant $k$ and effective mass $m$). The two resonators are coupled through a third spring, with spring constant $K = \epsilon k$. A sketch of the model is shown in Fig~\ref{fig:S6}. The two resonators are damped to mimic the experimental resonance quality factor. The torsion spring alone is actuated directly by a pure sinusoidal torque with magnitude $\tau_{\textrm{D}}$, set to a value characteristic of the maximum torque during the experimental hysteresis measurements.  The following coupled differential equations then describe the displacement of each spring from its equilibrium position, with $x_{1}$ representing the torsion spring and $x_{2}$ the linear spring:
\begin{equation}
\begin{split}
I\frac{d^{2}\theta}{dt^{2}} & =-(\kappa\theta + Kx_{1}R_\textrm{eff})+KR_\textrm{eff}x_{2}-\gamma_{\textrm{T}}\frac{d\theta}{dt}+\tau_{\textrm{D}}\cos(\omega t)\\ m\frac{d^{2}x_{2}}{dt^{2}} & =-(k+K)x_{2}-\gamma_{\textrm{L}}\frac{dx_{2}}{dt}Kx_{1},
\end{split}
\end{equation}
\noindent The first of the two equations describes the torsional motion coupled to the spring system while the second equation describes translational motion due to the linear spring.  The two oscillatory solutions represent symmetric and antisymmetric motions of $x_1$ and $x_2$.  Representative solutions plotted in Fig.~\ref{fig:S6}b) and c) show that $x_1$ and $x_2$ have developed similar peak amplitudes already when the coupling parameter is $\epsilon = 0.04$.  The relatively small frequency redshifts of the lower mode quickly vanish as $\epsilon$ increases and the symmetric motion ceases to stretch the coupling spring.  At the same time, the higher mode continues to blueshift on account of the increase in stiffness against antisymmetric motion as the coupling spring becomes less compliant.       

The experimental identification of the primarily torsion-like and primarily flexion-like modes is from the displacement patterns observed in raster-scanned images of the interferometric signal (thermomechanical and RF torque-driven).  The flexion-like mode is found to have lower frequency than the torsion mode in the present work.  This torsion mode is 5$\%$ higher in frequency, yielding an upper bound estimate for the coupling parameter of $\epsilon=0.025$. 

\subsection{Thermomechanical Calibration}

\begin{figure}[htb!]
	\centering
	\includegraphics{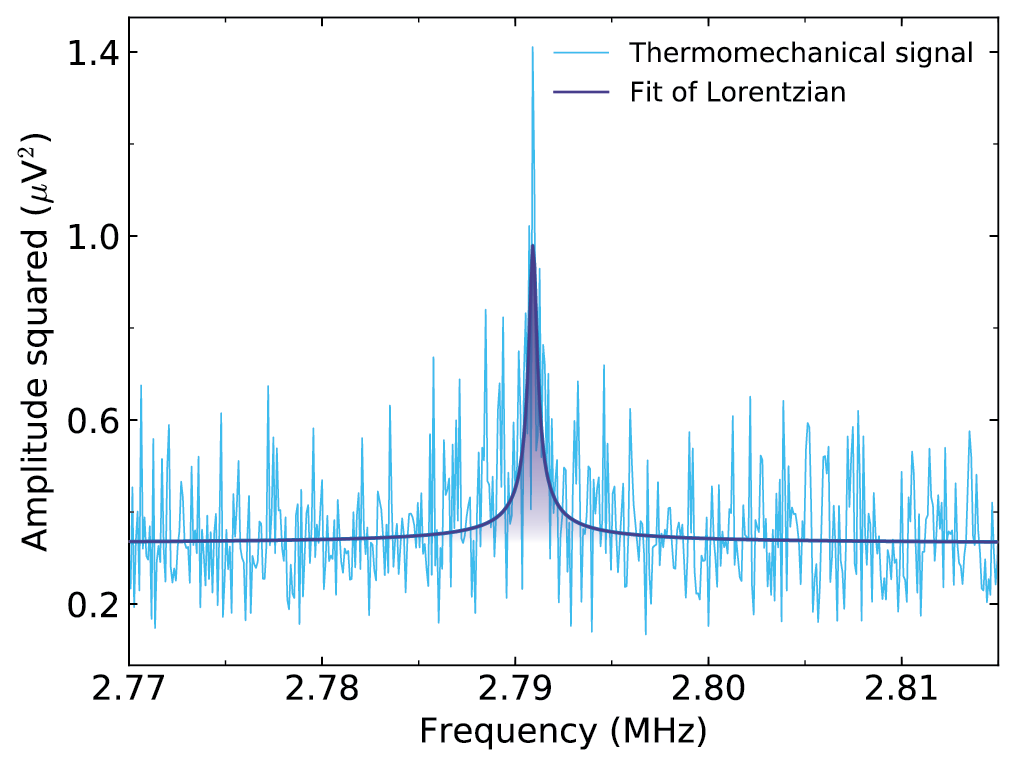}
	\caption{\label{fig:S7}%
	Squared thermomechanical (undriven) displacement signal recorded with a 46 Hz noise-equivalent power bandwidth.  The solid line is a Lorentzian fit.  The shaded peak area above the technical noise floor calibrates the torque response.}
\end{figure}

The total energies of the modes are given by

\begin{equation}
E_\textrm{tot} = \frac{1}{2}\kappa\Bigg(\frac{x_1^\textrm{peak}}{R_\textrm{eff}}\Bigg)^2 + \frac{1}{2}k(x_2^\textrm{peak})^2 + \frac{1}{2}K(x_1^\textrm{peak} \pm x_2^\textrm{peak})^2,
\end{equation}

\noindent where $`-$' and $`+$' in the third term correspond to the symmetric and antisymmetric modes.  The thermomechanical calibration is obtained by parsing the $\frac{1}{2}k_\textrm{B}T$ thermal energy between the $\kappa, k, \textrm{and}\thinspace K$ contributions.

The model parameters as obtained from a COMSOL model using device dimensions measured from an electron micrograph are $m = 1.79 \times 10^{-13}\thinspace$kg, $I = 4.78 \times 10^{-26}\thinspace$kg m$^2$, and $\kappa = 1.7 \times 10^{-13}\thinspace$Nm/rad.  To estimate the upper bound on $\epsilon$ it is assumed that the two modes are nearly degenerate when uncoupled (a 0.5\% initial frequency separation is chosen for clarity of the plots in Fig.~\ref{fig:S6}).  The thermomechanical signal in Fig.~\ref{fig:S7} then yields a peak Brownian displacement of $2.2 \times 10^{-11}\thinspace$m.  The final result is close to the value obtained by a simpler calibration treating the resonance as a pure torsion mode \cite{Losby2012}.  The loss of transduction efficiency from translational motion also being driven by torque is partially offset by selecting the measurement location where the two displacement contributions add.  The corresponding equivalent torque to drive the thermomechanical optical signal amplitude of 120 nV/Hz$^{1/2}$ is $\tau = 6.0 \pm 0.5$ zN$\cdot$m.  

\bibliography{RF_EdH_refs-v5-30apr2020.bib} 